\documentclass[twocolumn, usenatbib, fleqn]{mn2e} 
\usepackage{times}
\usepackage{graphicx}
\usepackage{epsfig}
\usepackage{amsfonts}
\usepackage{amsmath}
\usepackage{amssymb}
\usepackage{rotating} 
\usepackage{setspace}

\DeclareMathAlphabet{\mathbf}{OML}{cmm}{b}{it}
\DeclareMathAlphabet{\mathbfsf}{OT1}{cmss}{bx}{n}

\newcommand\bq{\begin{equation}}
\newcommand\eq{\end{equation}}

\newcommand\bqa{\begin{eqnarray}}
\newcommand\eqa{\end{eqnarray}}
\newcommand\bqas{\begin{eqnarray*}}
\newcommand\eqas{\end{eqnarray*}}

\newcommand\lra{\longrightarrow}
\newcommand\ra{\rightarrow}

\def\apj{{ ApJ}}
\def\apjs{{ ApJS}}
\def\apjl{{ ApJ}}
\def\aap{{ A\&A}}

\def\mnras{{ MNRAS}}

\def\araa{{ ARAA}}
\def\beq{\begin{equation}}
\def\eeq{\end{equation}}
\title[Radiative transfer of energetic photons: X-rays and helium ionization in \textsc{C}$^2$-\textsc{Ray}] 
{Radiative transfer of energetic photons: X-rays and helium ionization in \textsc{C}$^2$-\textsc{Ray}}
\author[Friedrich et al.]{Martina M. Friedrich$^{1}$
\thanks{e--mail: martina@astro.su.se}, 
Garrelt Mellema$^{1}$,
Ilian T. Iliev$^{2}$ and
Paul R. Shapiro$^{3}$\\
$^1$Department of Astronomy \& Oskar Klein Centre, AlbaNova, Stockholm University, SE-106 91
Stockholm, Sweden \\
$^2$Astronomy Centre, Department of Physics \& Astronomy, Pevensey II Building, 
University of Sussex, Falmer, Brighton BN1 9QH\\
$^3$Department of Astronomy and the Texas Cosmology Center, The University of 
Texas at Austin, TX 78712, USA\\
}
\date{\today}
\pubyear{2010} \volume{001} \pagerange{1}
\begin{document}
\pagerange{\pageref{firstpage}--\pageref{lastpage}}
\maketitle
\label{firstpage}
\begin{abstract}
We present an extension to the short-characteristic ray-tracing and non-equilibrium photon-ionization code \textsc{C}$^2$\textsc{-Ray}. The new version includes the effects of helium and improved multi-frequency heating. The motivation for this work is to be able to deal with harder ionizing spectra, such as for example from quasar-like sources during cosmic reionization. We review the basic algorithmic ingredients of \textsc{C}$^2$\textsc{-Ray} before describing the changes implemented, which include a treatment of the full on the spot (OTS) approximation, secondary ionization, and multi-frequency photo-ionization and heating. We performed a series of tests against equilibrium solutions from \textsc{CLOUDY} as well as comparisons to the hydrogen only solutions by \textsc{C}$^2$\textsc{-Ray} in the extensive code comparison in \citet{comparison1}. We show that the full, coupled OTS approximation is more accurate than the simplified, uncoupled one. We find that also with helium and a multi-frequency set up, long timesteps (up to $\sim 10$\% of the recombination time) still give accurate results for the ionization fractions. On the other hand, accurate results for the temperature set strong constrains on the timestep. The details of these constraints depend however on the optical depth of the cells. We use the new version of the code to confirm that the assumption made in many reionization simulations, namely that helium is singly ionized everywhere were hydrogen is, is indeed valid when the sources have stellar-like spectra. 
\end{abstract}
\section{Introduction} 

Photo-ionization is one of the major radiative feedback processes in
astrophysics. The extreme ultraviolet (EUV) photons produced by massive stars in star
formation regions are capable of heating the gas to temperatures
around $10^4$~K and the heated ions produce copious amounts of
collisionally excited line radiation, leading to the well-known and
sometimes spectacular images of H~II regions \citep[such as for example
the Orion nebula,][]{2001ARA&A..39...99O}. Accreting black holes and
neutron stars also produce ionizing radiation and, depending on their
mass, can ionize smaller or larger regions around themselves. On
galactic scales, the emission from supermassive central black holes
are observed to produce ionization cones stretching into the galaxy's
immediate environment \citep[see e.g.][]{1989ApJ...345..730P}. Even at
the largest scales photo-ionization is important. Some time before
redshift 6, ionizing radiation escaped from the first generations of
galaxies and percolated through the intergalactic medium (IGM),
changing it from cold and neutral to warm and ionized. This process is
known as reionization and was the Universe's last global phase
transition \citep[see for example the review
in][]{2006Sci...313..931B}.

Traditionally photo-ionization codes concentrated on solving
equilibrium situations, as for example \textsc{CLOUDY}
\citep{1998PASP..110..761F}, \textsc{MAPPINGS}
\citep{1993ApJS...88..253S} and the three-dimensional code
\textsc{MOCASSIN} \citep{2008ApJS..175..534E} do. The main aim of
these codes is to accurately calculate line strengths for comparison
to spectroscopic observations.  However, since the increase in
pressure can drive powerful flows in the gas, there is also a need to
couple photo-ionization calculations to hydrodynamics. This
necessitates a simpler version of the radiative transfer, since it has
to be calculated in step with the hydrodynamics and for dynamic
calculations the individual line strengths are less interesting. The
history of these types of calculations goes back quite far, see
for example \citet{1986ARA&A..24...49Y} for an overview of
the work done before 1986.

For some of the applications mentioned above, a lower dimensional
approach (one or two dimensional) is sufficient. However, other
applications, such as cosmic reionization, require the transport to be
performed in the full three dimensions. Because of the higher
dimensionality here as well a simpler version of the radiation and
photo-ionization physics is typically implemented, although the level
of sophistication varies between methods \citep[see e.g.\ the first
generation methods of][]{1999MNRAS.309..287R, 2001NewA....6..359S,
  2001MNRAS.321..593N, 2001MNRAS.324..381C}. In some cases the
coupling to the dynamics was also done for cosmological applications
\citep[e.g.,][]{2002ApJ...575...33R, 2008ApJ...689L..81T,
  2008ApJ...685...40W}.

One of the simplifications which is often employed is to consider
hydrogen as the only element being photo-ionized. Since close to 10\%
of the gas is helium, this approximation is crude. However, as shown
for example in figure~2.4 of \citet{2006agna.book.....O}, for
typical O-star spectra the ionization of helium follows largely that
of hydrogen. So, if one is only interested in the structure of the
ionized regions, assuming that helium follows hydrogen is not a bad
approximation.

However, as one moves to harder spectra, this assumption becomes less
and less valid. Not only does one need to take into account that helium
can become doubly ionized, also the fact that the cross section of the
two types of helium start contributing substantially to the opacity of
the gas becomes an issue. This problem becomes especially important
when the ionizing spectrum is powerlaw-like, as one expects from hot
accretion disks around black holes. Specifically for the case of
cosmic reionization, where there is a possible contribution
from powerlaw-like sources such as quasars and mini-quasars, any
proper study of their contribution should consider both hydrogen and helium.

This has motivated us to extend the capabilities of the code
\textsc{C}$^2$-\textsc{Ray} to include the effects of helium.
\textsc{C}$^2$-\textsc{Ray} is a photon-conserving radiative transfer
code that uses short characteristic ray tracing and is described in
detail in \citet{methodpaper} (hereafter M+06) . It has been used extensively for
reionization simulations \citep[e.g.][ to name a few]{mellema06,
  2007MNRAS.376..534I, 2008MNRAS.391...63I}. It was also combined with
a hydrodynamics code \citep[\textsc{Capreole-C}$^2$,
][]{2006ApJ...647..397M}, and an ideal magnetohydrodynamics code
\citep[\textsc{Phab-C}$^2$, ][]{2006A&A...449.1061D,
  2011MNRAS.414.1747A}, to investigate galactic H~II
regions. Furthermore, it was tested against other non-equilibrium radiative transfer codes in
\citet{comparison1} (hereafter I+06) and in conjunction with a grid-based hydrodynamic
code \citep[for details, see][]{2006ApJ...647..397M} in a second
comparison project, including gas dynamics
\citep{2009MNRAS.400.1283I}. 

Adding helium implies introducing another source of
frequency-dependent opacity, thus making a multi-frequency approach
inevitable. In addition the on-the-spot approximation becomes more
complicated as one has to take into account how recombination photons
from helium affect the hydrogen ionization. As one moves to higher
photon energies, one should also take into account the secondary
ionizations caused by the superthermal electrons produced when high
energy photons ionize the atoms and ions. Including EUV and soft X-ray (SX) photons
therefore is a non-trivial extension of the photo-ionization
calculations which we describe and test in this paper.

In terms of the physical processes included the method we present
here is similar to a number of others published in recent years, but
differs in the algorithms used. The codes \textsc{CRASH}
\citep{2009MNRAS.393..171M} and \textsc{LICORICE} \citep{2010A&A...523A...4B}
use Monte Carlo techniques. The codes \textsc{SPHRAY}
\citep{2008MNRAS.386.1931A} and \textsc{TRAPHIC} \citep{2011MNRAS.412.1943P}
implement ray tracing on particle data whereas \textsc{RADAMESH}
\citep{2011MNRAS.411.1678C} uses an adaptive mesh approach.

The lay-out of the paper is as follows. In Section
\ref{section:c2ray_basic} we give an overview of the basic algorithmic
ideas behind \textsc{C}$^2$-\textsc{Ray} to then proceed in Section
\ref{section:c2ray_extend} with an overview of how we extended its
capabilities to handle harder photons. Section \ref{section:tests}
contains the description of a series of one and three-dimensional
tests for the new method, also evaluating the effects of the various new
elements such as secondary ionizations and the coupled on-the-spot
approximation. A series of appendices describe several important elements
in more detail.

\section{Reminder of basic steps of the original \textsc{C}$^2$-\textsc{Ray} algorithm}
\label{section:c2ray_basic}
The \textsc{C}$^2$-\textsc{Ray} method was developed to be a time-dependent photo-ionization algorithm that could be efficiently combined with a hydrodynamics calculation, and not impose impractically short timesteps and small cell sizes on the latter. This is achieved by assuming that the ionization evolution of individual cells follows an exponential decay to the equilibrium solution and that a time-averaged value of the optical depth can be used to describe the effect of a cell on the radiative transfer during the entire timestep. This approach is able to correctly track the progress of ionization fronts over many cells during one timestep. In addition, optically thick cells are dealt with by defining the photo-ionization rate such that it is consistent with the number of photons absorbed inside a cell. \textsc{C}$^2$-\textsc{Ray} was described and tested in detail in M+06. Here we summarize some of the ideas in order to define our notation and provide an introduction to the extensions described in Sect.~\ref{section:c2ray_extend}.

The evolution of the ionized hydrogen fraction derives from the set of chemical evolution equations:
\bq
\frac{d}{dt} \left( \begin{array}{c} x_{\rm {HI}}\\
    x_{\rm {HII}} \end{array}
\right) = \left( \begin{array}{c} -
    \Gamma_{\rm {HI}}+C_{\rm {HI}} n_e \quad 
    \phantom{-}\alpha_{\rm HII}^B n_e \\
    \phantom{-}\Gamma_{\rm {HI}}+C_{\rm {HI}} n_e \quad -\alpha_{\rm
      HII}^B n_e \end{array} \right) \,
\left( \begin{array}{c} x_{\rm {HI}}\\
    x_{\rm {HII}} \end{array} \right), 
\label{H_matrix}
\eq 
where $x_{\rm {HI}}$ is the neutral hydrogen fraction, $x_{\rm{HII}}$ is the ionized hydrogen fraction, $n_e$ is the electron density, $\Gamma_{\rm {HI}}$ is the hydrogen photo-ionization rate (see below), $C_\mathrm{HI}$ is the collisional ionization rate and $\alpha_{\rm HII}^B$ is the recombination rate. Since $x_{\rm {HI}}+x_{\rm{HII}}=1$, we obtain
\bq
\frac{d}{dt} x_{\rm {HII}} =  \underbrace{-(\Gamma_{\rm {HI}}+C_{\rm {HI}} n_e  +\alpha_{\rm HII}^B n_e )}_{\mathsf{A}^{\rm H}} \, x_{\rm {HII}} + \underbrace{(\Gamma_{\rm {HI}}+C_{\rm {HI}} n_e)}_{g^{\rm H}}\, .
\label{H_equation}
\eq
Here we introduce the notation, $\mathsf{A}^{\rm H}$ and $g^{\rm H}$, which allows us to write the equation in the general vector form, useful later on,
\bqa
\frac{d}{dt} \, \mathbf{x} = \mathsf{A} \, \mathbf{x} + \mathbf{g}\,.
\label{general_problem}
\eqa
As is well known, the general solution $\mathbf{x}(t)$ to a set of equations of this type is the sum of the solution to the homogeneous case, $\mathbf{x}_h(t)$, where $\mathbf{g}=\mathbf{0}$, and a particular solution $\mathbf{x}_p$: 
\bqa 
\mathbf{x}(t)= \mathbf{x}_h(t)+\mathbf{x}_p \quad \quad \textrm{with } \quad \mathbf{x}_h=\sum_{i=1}^n c_i \, \mathbfsf{x}_i \, \mathrm{e}^{t \lambda_i} \, .
\label{general_solution}
\eqa
Here, $n$ is the rank of $\mathsf{A}$, i.e.\ the number of coupled equations (1 in the case of hydrogen), $\lambda_i$ are the eigenvalues of $\mathsf{A}$, $\mathbfsf{x}_i$ are the corresponding eigenvectors and $c_i$ are coefficients which can be calculated from the boundary condition $\mathbf{x}(t=0)=\mathbf{x}_0$: 
\bqa
\mathbf{x}_0=\sum_{i=1}^n c_i \, \mathbfsf{x}_i + \mathbf{x}_p
\label{find_coeffs}
\eqa 
In subsequent timesteps, $\mathbf{x}_0$ is the state at the end of the previous timestep.  In the case of a constant $\mathbf{g}$, the particular solution can be the equilibrium solution given by
\bqas
\mathsf{A} \, \mathbf{x_p} + \mathbf{g} =   \mathbf{0} \,.
\eqas
For the simplest, hydrogen only case, the coefficients are thus
\begin{eqnarray}
  \lambda^H&=&-(\Gamma_{\rm {HI}}+C_{\rm {HI}} n_e  +\alpha_{\rm HII}^B n_e)\nonumber\\
  \mathbfsf{x}^H&=&1\nonumber\\
  x_p^H&=&\frac{\Gamma_{\rm {HI}}+C_{\rm {HI}} n_e}{\Gamma_{\rm {HI}}+C_{\rm {HI}} n_e  
    +\alpha_{\rm HII}^B n_e}\nonumber   \\
  c^H&=&x_0-x_p^H
  \label{Eq:H_values}
\end{eqnarray}
Here, we added the superscript $H$ for hydrogen and we skipped the subscript $1$ since for hydrogen, $n=1$.   

\begin{figure}
\centering
\includegraphics[width=8cm]{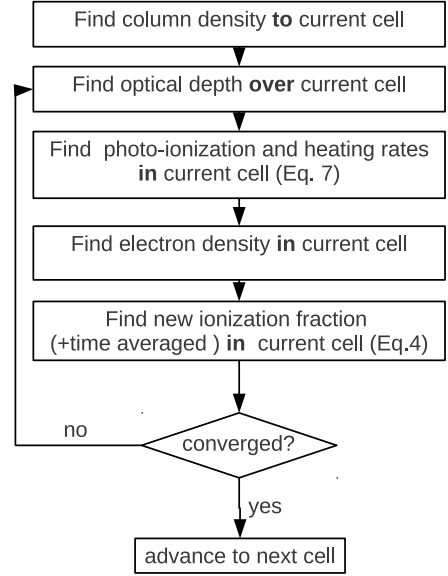}
\caption{Iteration scheme of \textsc{C}$^2$-\textsc{Ray}
    for a single cell, conceptionally as in M+06, figure 2}
\label{onecell_C2Ray_it}
\end{figure}

The photon-conserving photo-ionization rate $\Gamma$ in each cell used in $\mathsf{A}$ is calculated using 
\bqa
  \Gamma_{\rm {HI}}=\int_{\nu_{\mathrm th}}^\infty {L_\nu e^{-\langle
        \tau_\nu \rangle} \over h\nu}
        {1-e^{-\langle\Delta\tau_\nu\rangle}\over \langle n_{\mathrm HI}
        \rangle V_{\mathrm {shell}}}{\mathrm d}\nu\,,
\label{photoncons_gamma}
\eqa
where $\langle\Delta\tau_\nu\rangle$  is the time averaged optical depth over the cell and $V_{\mathrm {shell}}$ is the volume of the shell the cell belongs to. This quantity can be calculated from the time evolution of the neutral fraction  (Eq.~\ref{general_solution}). By solving these two equations (\ref{general_solution} and \ref{photoncons_gamma}) in alternating order one iterates to convergence as illustrated in Fig.~\ref{onecell_C2Ray_it}. This iteration also involves the electron density $n_e$ which is calculated from the time averaged ionized fraction. The time averaged optical depth to the cell $\langle\tau_\nu\rangle$ is calculated by short characteristic ray-tracing over the solutions found for cells lying nearer to the source. This makes the algorithm causal.

In the case of multiple sources, the iteration as shown in Fig.~\ref{onecell_C2Ray_it} is split up in two parts: The first part, including step three (finding the ionization and heating rates in each cell) is done for each source, looping through the entire computational grid using short characteristic ray tracing. For each cell, the rates from all sources are added. These total rates are used in the remaining two steps in the iteration. See also M+06 for a flow chart and a description of the implementation how to loop through the source list.

Note however that the flow chart in M+06 for the single source loop (figure 4) incorrectly includes the last two steps of the single cell loop. The electron density and the time averaged ionization fractions are in fact not updated in the source loop but are updated first after the photo-ionization rates from all sources are summed to a global photo-ionization rate. 

\section{Extending \textsc{C}$^2$-\textsc{Ray}}
\label{section:c2ray_extend}
The original \textsc{C}$^2$-\textsc{Ray} methodology works well for soft, stellar spectra. In this case, the IGM can be considered to be hydrogen only since there are not many photons capable of ionizing He~II and helium can be assumed to be singly ionized everywhere where hydrogen is ionized. In Sect. \ref{sec:cosmo_test} we show that the morphology of the ionization fraction field in a cosmological reionization simulation with stellar sources (only), is indeed hardly affected by the inclusion of helium. However, when the spectrum has a significant amount of SX photons, helium contributes significantly to the optical depth and a multi-frequency approach is required: at frequencies higher than the ionization threshold of $\rm {He II}$, the ionization cross-sections of $\rm {He   I}$ and $\rm {He II}$ are roughly an order of magnitude larger than the $\rm {H I}$ ionization cross-section. Therefore, neglecting helium in the case of sources with hard spectra will underestimate the optical depth substantially. We therefore have to add helium chemistry to \textsc{C}$^2$-\textsc{Ray}.

In order to include helium, both the chemical evolution equation, Eq.~(\ref{H_equation}), and the calculation of the ionization rate, Eq.~(\ref{photoncons_gamma}), have to be changed. Additionally, as shown below, the iteration scheme from Fig.~\ref{onecell_C2Ray_it} has to be modified. We describe each of these changes here.
 
\subsection{Chemical evolution equation} 
The procedure for adding helium photo-ionization to our calculations is by itself relatively straightforward as the basic algorithmic idea described in Sect.~\ref{section:c2ray_basic} provides the frame work for this. However, a complicating factor is the presence of ionizing recombination photons since they couple the rate equations of hydrogen and helium. Here we present two approaches for dealing with these, where the first one is less accurate, but simpler and more similar to what other authors have used. We compare the results of these two approaches in Sect.~\ref{section:tests}.
\subsubsection{Simple recombination: no coupling of species}
\label{no_coupling_eq}
When ions recombine, photons are emitted. In case of recombination of hydrogen ions, only recombinations to the ground state result in photons energetic enough to ionize hydrogen. If one assumes these photons to ionize immediately another hydrogen atom close by, this is called the \em on the spot \em (OTS) approximation for hydrogen \citep[e.g.][]{2006agna.book.....O}. In this approximation, the recombination coefficient to all states of hydrogen, $\alpha^A$ is replaced by the recombination coefficient to all states but the ground state, $\alpha^B$. For a mix of hydrogen and helium, the OTS approximation is more complicated as helium recombination photons can ionize both hydrogen and helium. However, as a first step, we assume that photons from recombinations to the ground state can only ionize the same species from which they originate and that in recombinations to other states than the ground state no ionizing photons are emitted. That means, we use the $\alpha^B$ recombination-coefficients for all species. In the following we refer to this as the ``uncoupled on-the-spot approximation'' (U-OTS). In this approximation hydrogen and helium can be treated separately. For helium, the set of chemical evolution equations (in analogy to Eq.~\ref{H_matrix} for hydrogen) is:
\begin{align}
 \frac{d}{dt} \left( \begin{array}{c} x_{\mathrm{HeI}}\\  
                                     x_{\mathrm{HeII}} \\
                                     x_{\mathrm{HeIII}} \end{array} 
             \right) =  \phantom{lllllllllllllllllllllllll llllllllllllllllllllllllllllllllll} \nonumber \\
          \left( \begin{array}{lcr} -U_{\mathrm{HeI}}\; 
                                                 &\phantom{-}n_e \alpha_{\mathrm{HeII}}\phantom{ -U_{\mathrm{HeII}}} 
                                                 &  0 \\
                                                 \phantom{-}U_{\mathrm{HeI}}\;
                                                 & \,-n_e \alpha_{\mathrm{HeII}} - U_{\mathrm{HeII}}  
                                                 &\phantom{-}  n_e \alpha_{\mathrm{HeIII}} \\
                                                 \phantom{-(}             0      \;   
                                                 & \; \phantom{-n_e \alpha_{\mathrm{HeII}} -}\, U_{\mathrm{HeII}}  
                                                 & - n_e \alpha_{\mathrm{HeIII}}
                       \end{array} \right) \, 
             \left( \begin{array}{c} x_{\mathrm{HeI}}\\
                                     x_{\mathrm{HeII}} \\
                                     x_{\mathrm{HeIII}} \end{array} 
             \right) \,
\label{He_matrix_form}
\end{align}
where $x_{\mathrm{HeI}}$ is the neutral helium fraction, $x_{\mathrm{HeII}}$ is the singly ionized helium fraction and $x_{\mathrm{HeIII}}$ is the doubly ionized helium fraction. We grouped the ionizing `up rates' into one term
\bqa
  U_n\equiv\Gamma_n+n_e C_n\,.
\eqa
The subscripts on $C$, $\alpha$ and $\Gamma$ indicate on which species they act, so $ \displaystyle \rm{He~II}
\mathop{\leftrightharpoons}^{\alpha_{\hbox{\tiny HeIII}}}_{\Gamma_{\hbox {\tiny HeII} } } \rm{He~III} $. 
Using the fact that $x_{\mathrm{HeI}}+x_{\mathrm{HeII}}+x_{\mathrm{HeIII}}=1$, the equivalent equation to Eq.~(\ref{H_equation}) can be written as 
\bqa
\frac{d}{dt} \left(\begin{array}{c}x_{\mathrm{HeII}} \\
    x_{\mathrm{HeIII}} \end{array} \right)
= \mathsf{A}^{\mathrm{He}} \cdot \left(\begin{array}{c}x_{\mathrm{HeII}} \\
    x_{\mathrm{HeIII}} \end{array} \right) +{\mathbf{g}^{\mathrm{He}}}\,,
\label{He_equation}
\eqa
where the vector $\mathbf{g}^{\mathrm{He}}$ and the matrix $\mathsf{A}^{\mathrm{He}}$ have the following forms
\begin{align}
   \mathbf{g}^{\rm He}&= \left(\begin{array}{c} U_{\mathrm{HeI}}\;\\
             0
             \end{array} \right) \nonumber \\
\mathsf{A}^{\rm He}&=\left( \begin{array}{cc}  \,-n_e \alpha_{\mathrm{HeII}} - U_{\mathrm{HeII}} -U_{\rm {HeI}}
                                      &\phantom{-}  n_e \alpha_{\mathrm{HeIII}}-U_{\rm {HeI}} \\
                                      \; \phantom{-n_e \alpha_{\mathrm{HeII}} -}\, U_{\mathrm{HeII}} 
                                      & - n_e \alpha_{\mathrm{HeIII}} \end{array} 
            \right) 
\end{align}
The solution of this set of linear differential equations for the U-OTS case can be found in \citet{2008MNRAS.386.1931A} or with the here introduced notation in Appendix \ref{solution_uncoupled}.
\subsubsection{On the spot approximation: coupling of species}
\label{ots_expl}
In reality, recombination photons from helium can ionize either hydrogen or helium, introducing the need to couple the rate equations for the two elements. This is the proper OTS approximation. Table \ref{recoms} gives an overview of the recombination processes affecting hydrogen and helium fractions in the OTS approximation.

To implement the OTS approximation we follow \cite{2006agna.book.....O} for dealing with the recombinations of $\mathrm{HeII}$ to $\mathrm{HeI}$ and \citet{1980MNRAS.191..301F} for those of $\mathrm{HeIII}$ to $\mathrm{HeII}$ (except for recombinations to the ground-state). Mostly we also use their notation.

For hydrogen, we take the $\alpha^B$-recombination coefficient $\alpha_{\mathrm{\rm {HII}}}^B$. For HeII, the photons from recombinations to the ground state are distributed between helium and hydrogen depending on the fraction of optical depth at the helium ionization threshold frequency, $\nu_{\rm{th}}^{\rm He I}$: a fraction $y$ goes into hydrogen ionization, a fraction $1-y$ goes into helium ionization. The photons from states other than the ground state contribute with a fraction $p$ to hydrogen ionization. Similarly, for HeIII the recombinations to the ground state ionize HeII, HeI and HI, depending on the relative optical depth over the cell in question at $\nu_{\rm{th}}^{\rm He II}$. Of those recombination photons, a fraction $y_2^a$ goes into HeII ionization, a fraction $y_2^b$ goes into HeI ionization and a fraction $1-y_2^a-y_2^b$ goes into HI ionization. Here, the fractions $y$, $y_2^a$, and $y_2^b$ are dependent on the relative optical depth of the species at the threshold frequencies of $\mathrm{HeI}$ and $\mathrm{HeII}$, as described below.

Recombinations of HeIII to other states than the ground state contribute to ionization of HI and HeI. Those recombinations lead either to two photon emission \citep[in a fraction $v$ of the cases, where $v$ is temperature dependent,][]{1964MNRAS.127..217H}, of which on average a fraction $l$ is energetic enough to ionize HI and a fraction $m$ is energetic enough to ionize HeI; therefore, a fraction $v \,w=v\,(l-m+m \,y)$ goes into HI ionization and a fraction  $v\,(m\,(1-y))$ goes into HeI ionization. The remaining fraction, $1-v$, leads to emission of a He Lyman $\alpha$ photon. Those photons are absorbed by any species to a fraction $f$. By letting escape some of the helium Ly $\alpha$ photons ($f \neq 1$) we lose them since we do not include those at larger distances from the source.  Of the absorbed He Lyman $\alpha$ photons, a fraction $z$ goes into HI ionization and the remaining fraction $1-z$ into HeI. Additionally, the Balmer continuum emission photons ($\alpha^2_{\rm{HeIII}}$) can ionize hydrogen. Table \ref{recoms} summarizes the photon emitting recombination processes included in the on the spot treatment. The numerical parameters used are listed in Table~\ref{table:ots_parameters}.

\begin{table*}
\caption{Summary of recombination processes included in the OTS treatment}
\begin{tabular}{|lclclcl|}
\hline
$\rm{HII}$ & $\stackrel{\alpha_{\rm {HII}}^1}\lra \rm{HI}$ & ground state recomb     & $\Rightarrow$ &&& $\rm{HI} \ra \rm{HII}$ ionization     \\

$\rm{HeII}$ & $  \stackrel{\alpha_{\rm HeII}^B} \lra \rm{HeI}$ & deexciations from  & $\stackrel{p}\Rightarrow$ &&& $\rm{HI} \ra \rm{HII}$ ionization \\
&&recombinations to n$\ge$ 2&&&&\\
$\rm{HeII}$ & $  \stackrel{\alpha_{\rm HeII}^1} \lra \rm{HeI}$ & ground state recomb    & $\stackrel{y}\Rightarrow$ &&& $\rm{HI} \ra \rm{HII}$ ionization  \\
       &            &                        & $\stackrel{1-y}\Rightarrow$ &&& $\rm{HeI} \ra \rm{HeII}$ ionization                 \\
$\rm{HeIII}$ & $\stackrel{\alpha_{\rm {HeIII}}^1}\lra \rm{HeII}$ & ground state recomb & $\stackrel{y_2^a}\Rightarrow$ &&& $\rm{HeII} \ra \rm{HeIII}$ ionization \\
&&&$\stackrel{y_2^b}\Rightarrow$ &&& $\rm{HeI} \ra \rm{HeII}$ ionization \\
&&&$\stackrel{1-y_2^a-y_2^b}\Rightarrow$ &&& $\rm{HI} \ra \rm{HII}$ ionization \\
$\rm{HeIII}$ & $ \stackrel{\alpha_{\rm {HeIII}}^2} \ra \rm{HeII}$ & recomb to n = 2 &  $\Rightarrow$ & HeII Balmer continuum & $\Rightarrow$ & $\rm{HI} \ra \rm{HII}$ ionization     \\
$\rm{HeIII}$ & $ \stackrel{\alpha_{\rm {HeIII}}^B} \ra \rm{HeII}$ & deexcitations from  & $\stackrel{v}\Rightarrow$ & 2 photon decay  & $\stackrel{w}\Rightarrow$&   $\rm{HI} \ra \rm{HII}$ ionization \\ 
&&recombinations to n $\ge$ 2&&& $\stackrel{m(1-y)}\Rightarrow$& $\rm{HeI} \ra \rm{HeII}$ ionization  \\
          &            &                        & $\stackrel{1-v}\Rightarrow$ & $\rm{HeII}$ Ly$\alpha$ photon   & $\stackrel{f(z-1)}\Rightarrow $ &$ \rm{HeI} \ra \rm{HeII}$ ionization   \\
         &      &&       & & $\stackrel{fz}\Rightarrow$ &$ \rm{HI} \ra \rm{HII}$ ionization \\ 

\hline
\label{recoms}
\end{tabular}
\end{table*}

\begin{table*}
\caption{Overview of the numerical parameters used in the OTS approximation}
\begin{center}
\line(1,0){450}
\end{center}
\bqas
p &=& \textrm{0.96 or 0.66 depending on $n_e$ \citep[][]{2006agna.book.....O}, in our cases of interest (cosmological simulations) always 0.96}   \\
y &=& \tau_{\rm H} /(\tau_{\rm H} + \tau_{\rm HeI}) \textrm{ at } \nu_{\rm{th}}^{\rm He I} \textrm{ (ionization threshold of He I) }\\
y_2^a &=& \tau_{\rm{HeII}} /(\tau_{\rm H} + \tau_{\rm{HeI}} +\tau_{\rm{HeII}})\; \rm{at}\; \nu_{\rm{th}}^{\rm He II} \textrm{ (ionization  threshold of He II)} \\
y_2^b &=& \tau_{\rm{HeI}} /(\tau_{\rm H} + \tau_{\rm{HeI}} +\tau_{\rm{HeII}})\; \rm{at}\; \nu_{\rm{th}}^{\rm He II}\textrm{ (ionization threshold of He II}) \\
z &=& \tau_{\rm H} /(\tau_{\rm H} +\tau_{\rm HeI}) \;\rm{at} \;h \nu \; = 40.8 \rm eV \, \textrm{(He I  Ly $\alpha$ )} \\
f &=& 1 \textrm{ to 0.1 ("escape" fraction  of Ly $\alpha$ photons) depending on the neutral fraction} \\
v &=& \textrm{temperature dependent coefficient}  \textrm{\citep[][]{1964MNRAS.127..217H}} \\
w &=& (l-m)+m \, y  \\
l &=& 1.425 \textrm{, fraction of photons from 2-photon decay, energetic enough to ionize hydrogen \citep[][]{1980MNRAS.191..301F}} \\
m &=& 0.737 \textrm{, fraction of photons from 2-photon decay,
  energetic enough to ionize neutral helium
  \citep[][]{1980MNRAS.191..301F}} 
\eqas 
\begin{center}
\line(1,0){450}
\end{center}
\label{table:ots_parameters}
\end{table*}

We can now introduce these terms in the general equation, Eq.~(\ref{general_problem}):
\bqa
\label{coupled_x_g}
\begin{array}{ccc}
     \mathbf{x} = \left( \begin{array}{c}
                                     x_{\rm {HII}} \\
                                     x_{\rm {HeII}} \\
                                     x_{\rm {HeIII}} 
               \end{array} \right) &
     \mathbf{g} = \left( \begin{array}{c} 
                                     U_{\rm {HI}} \\
                                     U_{\rm {HeI}} \\
                                     0
               \end{array} \right)
\end{array}
\eqa
and

\bq
\mathsf{A}=\left(\begin{smallmatrix}
                     &                                             & \frac{n_{\rm He}n_e}{n_{\rm H}} \times \\
-(U_{\rm HI}+        & \frac{n_{\rm He}n_e}{n_{\rm H}} \times      & ( (fz(1-v)+vw)\alpha^B_{\rm HeIII}+ \\
+\alpha_{\rm H}^Bn_e)& (y\alpha^1_{\rm HeII}+p\alpha^B_{\rm HeII}) & \alpha^2_{\rm HeIII}+(1-y^a_2-y^b_2)\alpha^1_{\rm
										         HeIII})) \vspace{3mm}\\
                     &                                                    & -U_{\rm HeI +} \\	
                     & -U_{\rm HeII}-U_{\rm HeI}-	         & n_e \left( \alpha^1_{\rm HeIII}y_2^b + \right. \\
0                    & (\alpha^A_{\rm HeII}-(1-y)\alpha^1_{\rm HeII}) n_e & (\alpha_{\rm HeIII}^A-\alpha^1_{\rm HeIII}
												    y^a_2) +          \\     
			   &                                     &   \left. \alpha^B_{\rm HeIII} (f(1-z)(1-v)+(l-w)v)\right)
			   							    \vspace{3mm}\\
0                    &      U_{\rm HeII}           &   (-n_e (\alpha_{\rm HeIII}^A-y_2^a \alpha^1_{\rm He III})) 
\end{smallmatrix} \right)
\label{coupled_A}
\eq

The density fractions $n_{\rm He}/n_{\rm H}$ in $\mathsf{A}_{12}$ and $\mathsf{A}_{13}$ are due to the fact that the equations are written in terms of fractions, not in terms of densities. The complete solution for this set of equations is presented in Appendix \ref{solution_coupled}.
\subsection{Extending the iteration mechanism}
\begin{figure}
\centering
\includegraphics[width=10cm]{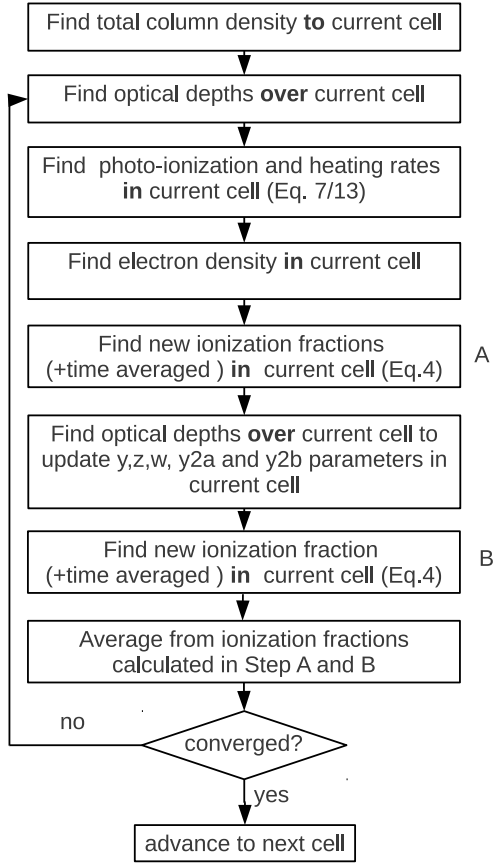}
\caption{Extended iteration scheme for \textsc{C}$^2$-\textsc{Ray} including helium with coupling of the species.}   
\label{new_iteriation_mechanism}
\end{figure}

In Sect. \ref{ots_expl}, we treat the set of chemical evolution equations $\mathrm{d}x/\mathrm{d}t=f(x)$ as a set of linear differential equations, Eq.~(\ref{general_problem}), and use iterations to obtain the correct electron density and ionization rates. However, further dependences on the different ionization fractions are hidden in the parameters $y$, $z$, $w$, $y_2^a$ and $y_2^b$. We found that this hidden extra non-linearity can complicate the convergence of the iteration. We therefore extended our iteration scheme to make it more robust. The extension consists of updating the parameters $y$, $z$, $w$, $y_2^a$ and $y_2^b$ after the chemical evolution equation has been solved and then solving the chemical evolution equation for a second time with this new set of parameters. We then take the mean of these first and second solutions for the time-averaged ionization fractions and use this mean to calculate the photo-ionization rates. These extra steps mean that the iteration scheme now consists of seven instead of four steps, see Fig.~\ref{new_iteriation_mechanism}.

\subsection{Calculating the photon rates}
\label{radiation}
For a pure hydrogen medium Eq.~(\ref{photoncons_gamma}) gives the photo-ionization rate, i.e. $\Gamma=\Gamma_{\rm {HI}}$ and $\tau=\tau_\mathrm{HI}$. In the case of a medium consisting of hydrogen and helium, this simple treatment can only be used for photons with frequencies below the ionization threshold for neutral helium, $\nu_{\mathrm{th}}^{\mathrm{HeI}}$. Above the ionization threshold frequency of ionized helium, $\nu_{\mathrm{th}}^{\mathrm{HeII}}$ all three species, $\mathrm{HI}$, $\mathrm{HeI}$ and $\mathrm{HeIII}$ contribute to the optical depth. In the frequency bin between those threshold frequencies, $\mathrm{HI}$ atoms and $\mathrm{HeI}$ atoms contribute to the total optical depth. Therefore, the minimum number of frequency bins to consider separately when calculating the photo ionization rates for each species is three, henceforth referred to as \em (frequency) bins\em. As we explain below, we use a single frequency dependence for all species in each bin. Since the the different frequency dependences of the ionization cross sections $\sigma$ of the species are very different in the bins, it is useful to further subdivide bin 2 and bin 3, henceforth referred to as \em (frequency) sub-bins. \em To refer to a particular sub-bin, we use the following notation: $n.m$ refers to sub-bin $m$ of bin $n$ ($n=2,3$). In Appendix \ref{cross_sections} we show our power-law fits to the cross-section data from \citet{1996ApJ...465..487V}.

\begin{figure*} 
\begin{center}
 \includegraphics[width=10cm]{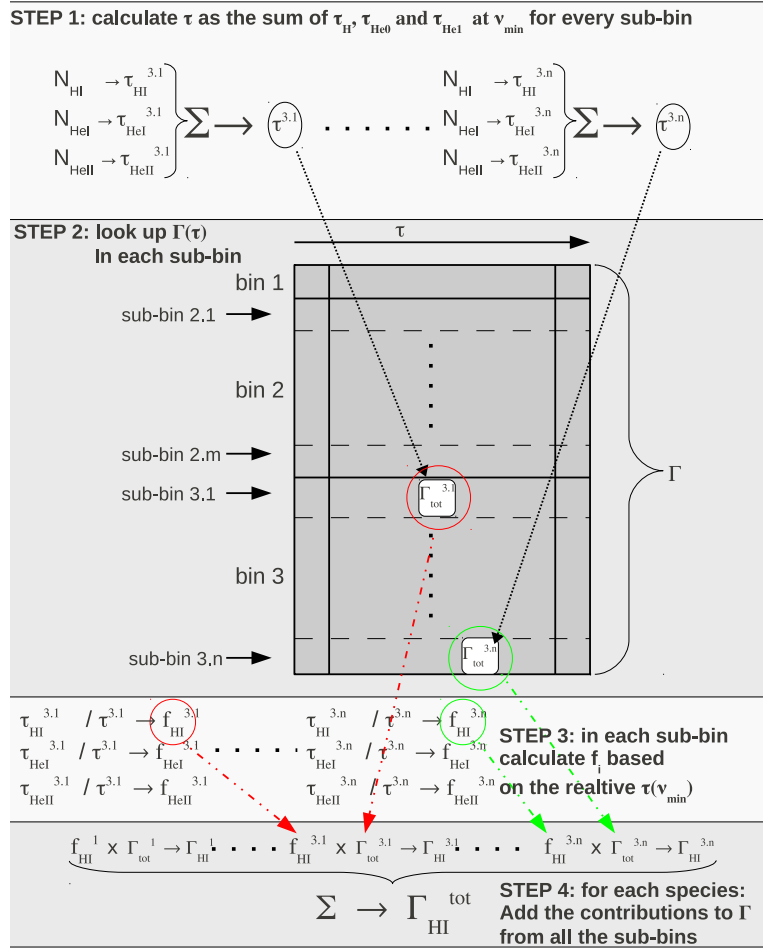}
\caption{Sketch showing how to calculate the ionization rate contribution for bin 3. In step 1, the total optical depth in each sub-bin is calculated. In step 2, the pre-calculated photo ionization rate table is used to calculate the total photo ionization rate in each sub-bin. On the basis of the optical depth of each species at the minimum frequency $\nu_{\mathrm{min}}$ in each sub-bin, the fractions $f_i$ going into hydrogen-, helium- and ionized helium- ionization are calculated in step 3. Finally, the products of the fractions and total photo ionization rates are summed over all sub-bins to result in the ionization rate for each species, $\Gamma_{\mathrm{H}}$, $\Gamma_{\mathrm{HeI}}$ and $\Gamma_{\mathrm{HeII}}$. 
\label{rad_diag}
}
\end{center}
\end{figure*}

The general treatment in every such sub-bin follows M+06 and is schematically illustrated in Fig.~\ref{rad_diag}. To avoid expensive integrations for each photo-ionization calculation, we tabulate the total ionization rate $\Gamma_{\mathrm{tot}}$ as a function of total optical depth $\tau_{\rm{tot}}$ at the minimum frequency of the sub-bin in question, assuming for all relevant species the same frequency dependence for the cross-section, following the concept outlined in \citet{1985MPARp.211....7T}. The assumption of having the same frequency dependence allows the summation of the optical depth of each species and the tabulation of the photo-ionization rate as a function of this single total optical depth. Since the frequency dependences of the cross-sections of the different species are in fact not identical (see for example Fig.~\ref{cross_secs2}), this is an approximation. The more sub-bins we use the more accurately we follow the actual shape of the (frequency dependence) curves of the different cross-sections. The frequency dependence imposed on all species is that of $\sigma_{\mathrm{HeI}}$ for all sub-bins of bin 2 and that of $\sigma_{\mathrm{HeII}}$ for all sub-bins of bin 3.  The table entries are the integrals of ionization rate over the frequency range of the sub-bin in question, thus the total ionization rate due to photons with frequencies in that frequency sub-bin.

In order to use this pre-calculated table and determine the total photo-ionization rate in a given cell, we need to determine the ionization cross-sections for every species at the minimum frequency of every sub-bin. We use the fitting formula from \cite{1996ApJ...465..487V} for the cross-sections and compute the total optical depth $\tau_{\mathrm tot}$ as the sum of the optical depths of all species at the minimum frequency of each sub-bin. This optical depth is used to determine the total photo ionization rate in the table.

Next, this total photo-ionization rate has to be split up into ionization rates for $\mathrm{HI}$ ($\Gamma_{\mathrm{HI}}$), $\mathrm{HeI}$ ($\Gamma_{\mathrm{HeI}}$) and in case of bin 3, $\mathrm{HeII}$ ($\Gamma_{\mathrm{HeII}}$). This is done according to the relative fractions of optical depth in each frequency interval.
\bq
\Gamma_i=\Gamma_{\mathrm{tot}} \frac{\tau_i}{\tau_{\mathrm{tot}}} \mathrm{,} \quad
\tau_{\mathrm{tot}}=\sum_i \tau_i \mathrm{,} \quad
\quad i=\mathrm{HI}, 
\mathrm{HeI} \textrm{ and } \, \mathrm{HeII}
\label{eq:rel_op_depth}
\eq
In Appendix \ref{photondivision} we evaluate other approaches for this distribution that have been proposed in the literature.

We calculate these fractions of optical depth at the minimum frequency of each sub-bin. This implies that we also here assume a single power law fit for the frequency dependence of all species in each sub-bin. Doing so overestimates the contribution of hydrogen to the optical depth in frequency bin 2 since the HI ionization cross section drops faster with frequency than the ionization cross section of HeI. This leads therefore to an overestimate of $\Gamma_{\mathrm{HI}} / \Gamma_{\rm{HeI}}$. In frequency bin 3, the slopes of the cross-section curves are in general more similar (see Appendix \ref{cross_sections}) but still steeper for $\mathrm{HI}$ which results in an overestimate of $\Gamma_{\mathrm{HI}} / \Gamma_{\mathrm{HeII}}$.  The more sub-bins used, the smaller the overestimates will be. We test the convergence of this in Appendix \ref{Subbins}. We find that increasing the number of sub-bins in bin 2 improves the slope of the ionization front while increasing the number of sub-bins in bin 3 improves the ionization fractions inside and outside the front.  

\subsection{Heating and secondary ionizations}
\label{sec:heating}

One of the motivations for extending \textsc{C}$^2$-\textsc{Ray}, is to investigate the effects that helium and hard spectra have on the temperature evolution of the IGM during the EoR.
In this section we first describe the implementation of the temperature calculation. This is followed by remarks on the inclusion of secondary ionizations and considerations about the additional timestep restrictions caused by the temperature calculation (in addition to the ionization calculation).

We use the ideal gas law $PV=N k_B T$, where $k_B$ is Boltzmann constant, to calculate the pressure $P$ from the temperature $T$ and the gas particle number density $N/V=n_{\rm H}+n_{\rm He}+n_e$ in each cell of volume $V$. 
From the pressure we then calculate the internal energy per volume $V$ (in each cell), using the dimensionless heat capacity (per particle) at constant volume, $c_V$: $u^{\rm{int}}=U^{\rm{int}}/V=c_V P$ 
For a monatomic gas, $c_V=3/2$. So the equation to convert temperature into internal energy density reads: 
\bq 
u^{\rm{int}}=\frac{3}{2} k_B T (n_{\rm H}+ n_{\rm He}+n_e)\, .
\eq
This energy density is affected by the heating ($\mathcal{H}$)  and cooling  ($\mathcal{C}$) of the gas per (cell-) volume and per unit time. In order to follow the temperature evolution of the gas, we therefore have to solve the equation
\bq
\frac{\partial u^{\rm{int}}}{\partial t}=\mathcal{H}-\mathcal{C} \,.
\label{eq:temperature}
\eq

As the only contribution to the heating rate $\mathcal{H}$, we consider photo-ionization heating. In every photo-ionization of species $i$, the excess energy, $(\nu - \nu_\mathrm{th}(i))$ is transferred to the electron released. If one assumes that the cells are optically thick, (almost) all photons are absorbed and the average energy per photo ionization is simply the average excess energy over the whole spectrum. This was the approach used by the original \textsc{C}$^2$-\textsc{Ray} in the tests with temperature evolution in I+06.\footnote{The H-only \textsc{C}$^2$\textsc{Ray} used in \citet{2006ApJ...647..397M} and \citet{2011MNRAS.414.1747A} actually used a one species, one frequency bin version of the heating method described below.}  

To properly take into account the effects of different optical depth on the heating we can calculate $\mathcal{H}$ in analogy to the ionization rate $\Gamma$ (Eq.~\ref{photoncons_gamma}). It should be remembered that, although we write this as one equation, for the photo-ionization rate it is in fact the difference between the ingoing photo-ionization rate (calculated on the basis of $\left<\tau\right>$) and the outgoing photo-ionization rate (calculated on the basis of the $\left<\tau\right>+ \left<\Delta \tau \right>$). For the heating, these quantities are rather abstract since they symbolize the  excess energy still in the form of photons when entering the cell and the amount of excess energy still in the form of photons leaving the cell. Nevertheless, what is tabulated as function of optical depth is this excess energy as a function of optical depth. The corresponding equation to  Eq.~(\ref{photoncons_gamma}) is then 
\begin{equation}
  \mathcal{H}(i)=\int_{\nu(\rm{sub-bin})} \frac{h(\nu - \nu_\mathrm{th}(i))L_\nu e^{-\langle
        \tau_\nu \rangle}}{h\nu}
  \frac{1-e^{-\langle\Delta\tau_\nu\rangle}}{V_{\rm shell}}{\rm d}\nu 
  \label{spatialgamma}
\end{equation}
for species $i$. Since the excess energy is different for each species, instead of having one table of heating rates in each sub-bin, we need three tables, where the excess energy is with respect to the threshold frequency of $\nu_{\rm th}({\rm H I})$, $\nu_{\rm th}({\rm He I})$ and $\nu_{\rm th}({\rm He II})$.
The total heating in each frequency sub-bin is naturally analogous to the photo-ionization given by 
\bq
\mathcal{H}=\sum_i \frac{\tau_i}{\tau_{\rm tot}} \mathcal{H}(i) \,.
\eq 

Some of the electrons that received excess energy $(\nu - \nu_\mathrm{th}(i))$ after being released from the bound state of atom $i$, will collide with bound electrons of other atoms (or ions)  rather than other free electrons. This transfers energy to the bound electron. If the transferred energy is greater than the binding energy of this electron, this electron is released. This process is called secondary ionization. The probability of such an event depends on the energy of the primary electron and on the ionization state of the gas. The full process requires careful modelling, but \citet{1985ApJ...298..268S}, \citet{2002ApJ...575...33R} and \citet{2008MNRAS.387L...8V}, to name a few, published separable functional relations dependent on primary electron energy and hydrogen ionization fraction to describe what fraction of the energy deposited  in primary electrons goes into secondary ionizations of either HI or HeI, $f_{\textrm{ion}}^{\textrm{HI/HeI}}$, and what into heat, $f_{\textrm{heat}}$. In these relations it is assumed that $x_{\rm{HII}}=x_{\rm{HeII}}$ and secondary ionizations of HeII are neglected. We implement the separable functional relationship as in \citet{2002ApJ...575...33R} which converges for high electron energies to the functional form of \citet{1985ApJ...298..268S}. 

\citet{2010MNRAS.404.1869F} pointed out that there is in general no simple separable functional form for deciding the fractions that go into secondary ionizations and heating. Rather the dependence of the relative fractions varies with ionization fraction in a complex way, see their figure 5. However, the complex relation given by \citet{2010MNRAS.404.1869F} still assumes $x_{\rm{HII}}=x_{\rm{HeII}}$ and no secondary ionizations of HeII. Given these limitations we decided that at this point there is no clear benefit in implementing these more computationally expensive relations. 

For the cooling rate, we include free-free and recombination cooling for H II, He II and He III and collisional excitation cooling for H I, He I and He II. A full overview of the radiative cooling rates used is given in Appendix~\ref{cooling_recom_rates}. In case of cosmological simulations, cosmological cooling due to the expansion of the universe and Compton cooling against the cosmic microwave background photons are included as well. 

In order to numerically solve Eq.~(\ref{eq:temperature}) we use forward Euler integration. However, the cooling rate depends sensitively on the gas temperature which is changing because f the heating. To accurately follow this behaviour we use the forward Euler method with sub-timesteps determined by a limit on the temperature change (typically 10\%). We keep the heating term $\mathcal{H}$ (which represents the average heating rate over the timestep) constant but use the temperature from the previous sub-timestep to calculate the cooling rate $\mathcal{C}$.

Above, we described how to calculate the heating rate in analogy to the photo-ionization rate. Specifically, we use the time-averaged optical depth (to the cell and in the cell) both in Eq.~(\ref{photoncons_gamma}) and Eq.~(\ref{spatialgamma}). There is however an important difference between the dependence of the photo-ionization rates and the heating rates on optical depth. Although Eq.~(\ref{photoncons_gamma}) gives the correct rate of absorbed photons, for the heating the frequency of the photons matters. Let us first consider the source cell, i.e., the optical depth to the cell,$\tau$, is constant with time and equal 0. While for the photo ionization,
\bq
\Gamma( \left< \tau \right>_{\Delta t} )=   \frac{\int_0^{\Delta t}\Gamma \left(\tau(t') \right) dt'}{\Delta t} \, ,
\eq
for the heating in general
\bq
\mathcal{H}( \left< \tau \right>_{\Delta t} ) \leq   \frac{\int_0^{\Delta t}\mathcal{H} \left(\tau(t') \right) dt'}{\Delta t} \, ,
\eq
because of the $(\nu-\nu_{\rm th}(i))$ term in the integral for $\mathcal{H}$. Qualitatively one could say that in the beginning of the timestep the heating per photo-ionization is the optically thick value (average photon energy over the whole spectrum), for later times it shifts to lower values, the limit of which is given by the optically thin value (average photon energy over the spectrum weighted with the photo-ionization cross-section)

For this first cell where the only dependence is on the optical depth inside the cell, it would be possible to tabulate a correction factor based on the change of the optical depth during a timestep. However, in the more general case where the optical depth to the cell depends on the time varying optical depth of the other cells along the ray, a local fix (such as using an energy-average based on the change of optical depth in the cell or sub-timestepping on a cell-by-cell basis) to the heating rate is no longer possible. We explored different approaches for calculating accurate heating rates using large timesteps, but we were unable to find a general solution. 

In Appendix \ref{sec:time_step_dep} we investigate how the heating depends on the choice of the timestep. From these tests we conclude the following:
\begin{itemize}
\item If we are interested in time scales larger than the recombination time, we can still use large timesteps, as the initial heating is no longer dominating.
\item If the cooling time $t_{\rm cool} < \Delta t$ we can also use large timesteps, as the temperature is set by the equilibrium heating and cooling rates. This is the case for typical interstellar medium conditions.
\item If we are interested in time scales below the recombination time, an accurate value for the temperature requires timesteps of the order of the ionization time. This constraint becomes more strict if the cells are very optically thick. 
\end{itemize}

Given these conclusions it is difficult to provide a simple recipe for choosing the timestep. We therefore recommend testing for numerical convergence if accurate temperatures are required.

\section{Testing the code}

\label{section:tests}
To test the validity of the approximations we made for the sake of code efficiency, we performed a series of tests. First of all we validated the new version against the old version of \textsc{C}$^2$-\textsc{Ray} by setting the helium abundance to a very low value. This test showed that the new version has the same photon-conserving properties as the method presented in M+06. For testing the time dependent solution with helium, including the OTS approximation, secondary ionization and temperature evolution, we would need a fully validated time dependent photo-ionization code, which as far as we know is not publicly available. We were therefore forced, just as for example \citet{2010A&A...523A...4B} and \citet{2011MNRAS.412.1943P} to compare our results against the one-dimensional photo-ionization equilibrium code \textsc{CLOUDY} \citep[version 08.00, last described in ][]{1998PASP..110..761F}. Below we present the results of two sets of such one-dimensional tests, one set in which the source has a black body (BB) spectrum with an effective temperature of $T_{\textrm{eff}}$, the other for a source with power-law (PL) like spectrum, where the energy-distribution can be described as $L(\nu) \propto \nu^{-\beta}$. For the BB source we test various aspects of the calculation of the ionization fractions while keeping the temperature constant. For the PL case we also consider the temperature evolution. We chose this approach since the latter case having a wide energy range of hard photons, constitutes a more difficult test for the photo-heating. Table \ref{test:param} gives an overview of the parameters used for all the one-dimensional tests. To demonstrate the multi-dimensional and multi-source capabilities of the extended \textsc{C}$^2$-\textsc{Ray} we also present the results of two test problems using three-dimensional cosmological density fields.

\begin{table}
\caption{Parameters for testing the code.}
\begin{tabular}{ll}
\hline
$n_H$/cm$^{-3}$ & constant hydrogen number density\\
$n_{He}$/cm$^{-3}$ &  constant helium number density \\
$\dot{N}_{\gamma}$/s & rate of photons in the energy interval \\ &  $[13.6, 5441.6]$ eV\\ 
$T_{\mathrm {eff}}$/K & effective temperature of the black body source in \\ & case of BB source \\
$\beta $& power law index in case of PL source \\
$T_{\mathrm{ini}},{\mathbf{x}}$& initial temperature and initial ionization states, H II, \\ & He II and He III   \\
$\Delta r$& the cell size \\
$\Delta t$&  the timestep  \\
$(n_2/n_3 ) $ & number of sub-bins in frequency bins 2 and 3 \\
OTS/U-OTS/ $\alpha^{\mathrm{A}}$  &   assuming the  (U-)OTS-approximation, or using \\ & $\alpha^\mathrm{A}$ recombination rates\\
\hline
\label{test:param}
\end{tabular}
\end{table}

\subsection{Test-suite 1: Black body source}
The first set of tests considers the expansion of an ionized region produced by a single source into a constant density medium with constant temperature. Under the assumption of spherical symmetry this can be calculated in one dimension for the radial distance $r$ to the source. The parameters that are not varied in this set of tests are:
$n_\mathrm{H}$/cm$^{-3}=1.0 \times 10^{-3}$,
$n_{\mathrm{He}}$/cm$^{-3}= 8.70 \times 10^{-5}$,
$\dot{N}_{\gamma}$/s  $= 5.0 \times 10^{48}$,
$T_{\mathrm {eff}}$/K  $=10^{5}$.
$T_{\mathrm{ini}}=10000 \mathrm{K},
\mathbf{x} = (10^{-40},10^{-40},10^{-40})$ (i.e. completely neutral),
$\Delta r$/pc$=150$,
$\Delta t$/yr=$10^7$ and
$(n_2 , n_3) =(10,14)$.

We choose these parameters so that the helium fraction $f_{\mathrm{He}}=n_{\mathrm{He}}/(n_\mathrm{H}+n_{\mathrm{He}})= 0.08$ and the remaining physical parameters are the same as in \texttt{Test 2} of I+06, except that we do not follow the temperature evolution here. The timestep $\Delta t$ is $\sim 10$\% of the recombination time in a fully ionized medium, $n_\mathrm{H}\alpha^\mathrm{A,B}$, and $\sim 10^4$ times the ionization time for the first cell.

\subsubsection{Test 1 A}
\begin{figure*}
\begin{center}
  \includegraphics[width=15cm]{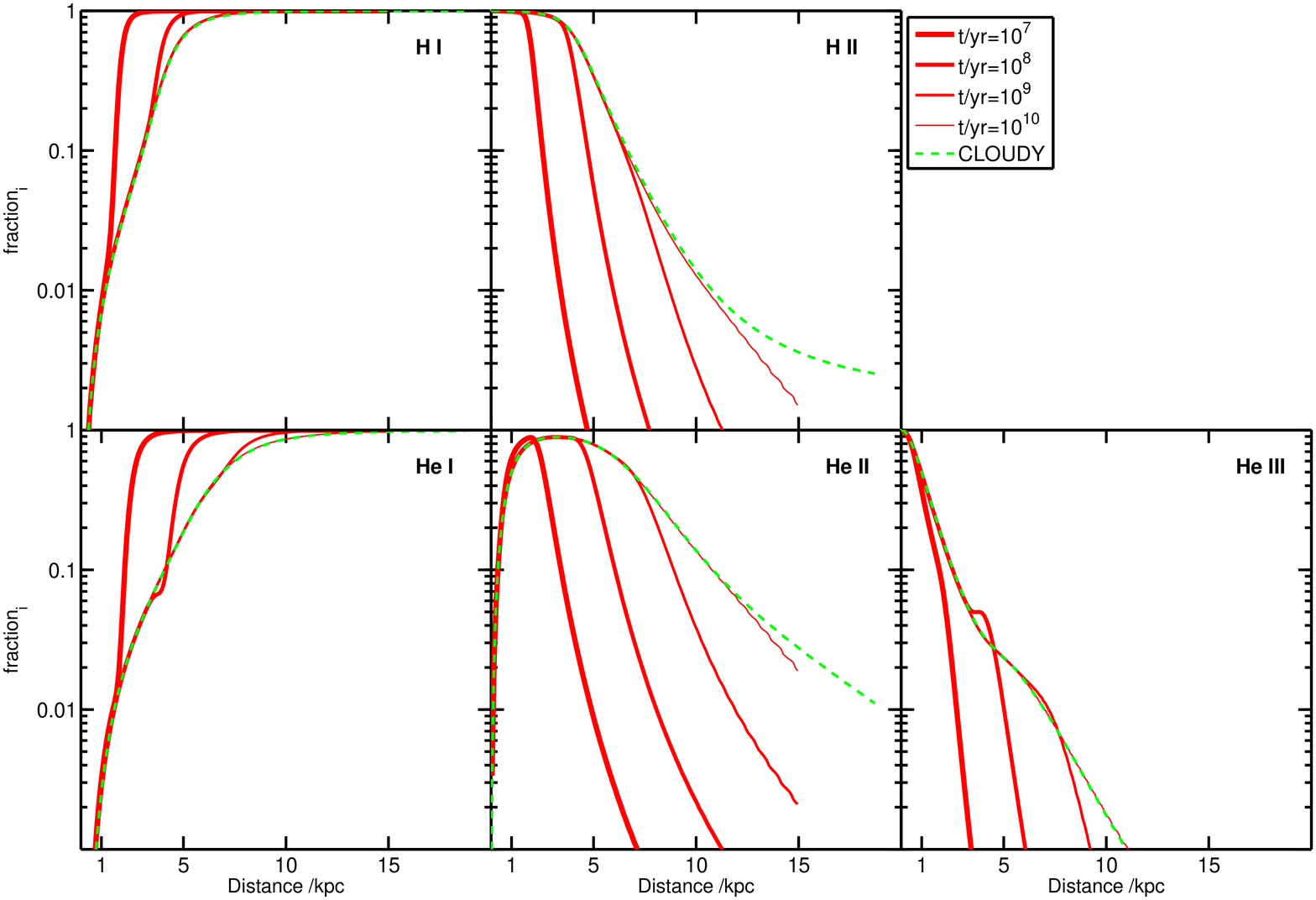}
\caption{Results from \texttt{TEST1 A}. Fractions of $\mathrm{H I}$ (upper left panel), $\mathrm{H II}$ (upper middle panel), $\mathrm{He I}$ (lower left panel), $\mathrm{He II}$ (lower middle panel) and $\mathrm{He III}$ (lower right panel) at times $t/ \mathrm{yr} =[1\times 10^7, 1 \times 10^8, 1 \times 10^9, 1 \times 10^{10}]$ as indicated by line thickness in the legend. We also show the equilibrium solution of \textsc{CLOUDY} (green dashed). Note that due to the low electron density in the outer regions, the equilibrium solution has not yet been reached beyond a distance of about 12 kpc at  $t=10^{10}$ yr.
\label{test1a_plot}
}
\end{center}
\end{figure*}

In the first test of this suite we assume that all ionizing photons from recombinations can escape, which means that we are using the $\alpha^{\mathrm{A}}$ rates. This test is meant to show how well our basic approach can match the equilibrium solution from \textsc{CLOUDY}. In Fig.~\ref{test1a_plot} we show the results together with that equilibrium solution.
It can be seen that after $t=10^{10}$ yr the general agreement with the \textsc{CLOUDY} equilibrium solution is excellent for distances below 12~kpc, but not for larger distances. However, since the recombination timescale is given by $(n_\mathrm{e}\alpha)^{-1}$, it follows that for ionization fractions below $\sim 1$\%, the recombination timescale is larger than $\sim 10^{10}$ yr. In other words, these outer regions have not yet reached equilibrium. Another noticeable feature in Fig.~\ref{test1a_plot} is that the $\mathrm{He III}$ fraction shows a transient bump around $r=4$~kpc. This bump disappears as the equilibrium solution is approached, but both, our results and the equilibrium solution still show a slope change in the $\mathrm{He III}$ curve around that same distance. From all this we conclude that the results of this test show that our basic multi-frequency radiative transfer is consistent with \textsc{CLOUDY}'s.

\subsubsection{Test 1 B}

\begin{figure*}
\begin{center}
  \includegraphics[width=15cm]{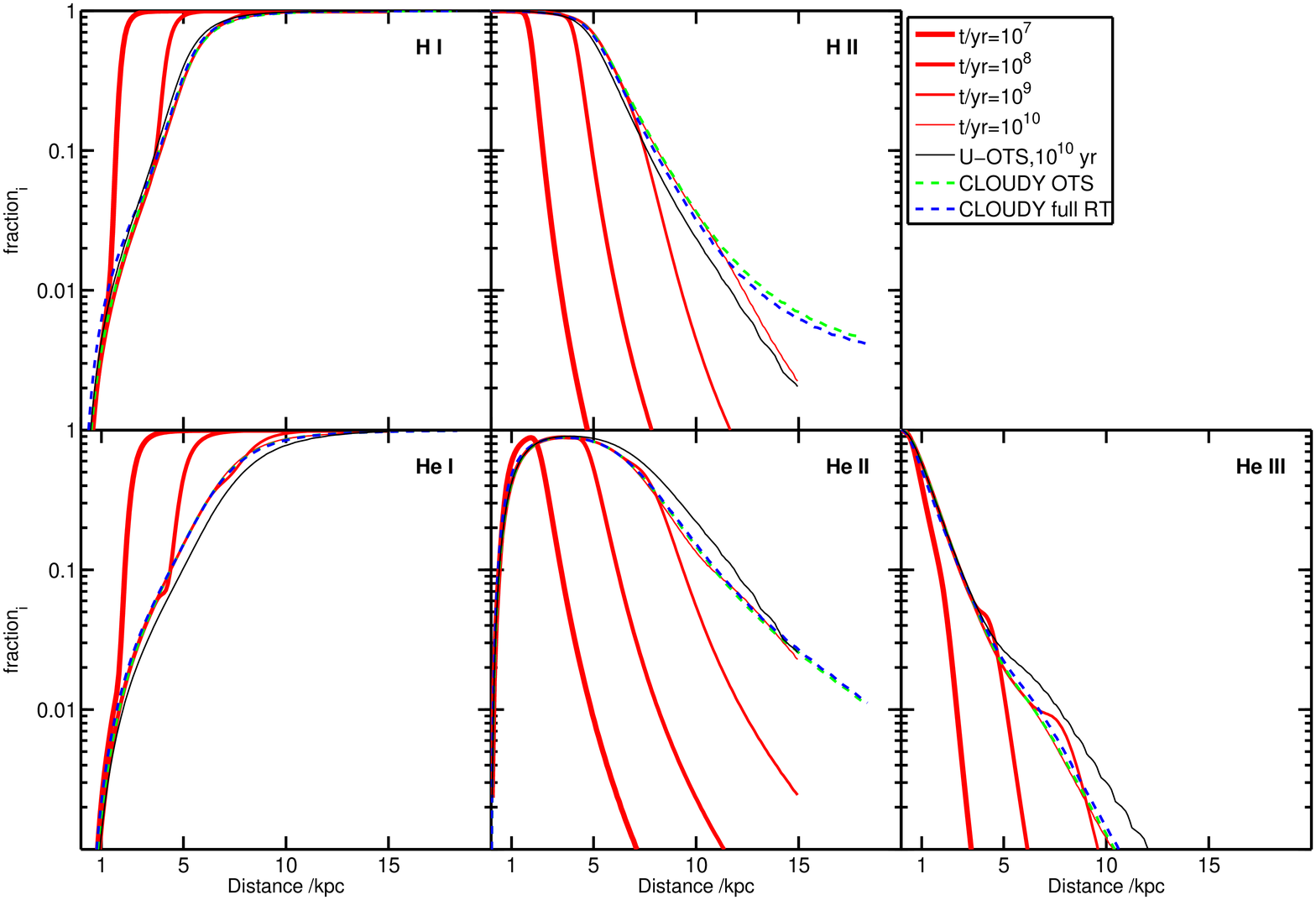}
\caption{Results from \texttt{TEST1 B}. Fractions of $\rm{H I}$ (upper left panel), $\rm{H II}$ (upper middle panel), $\rm{He I}$ (lower left panel), $\rm{He II}$ (lower middle panel) and $\rm{He III}$ (lower right panel) at times $t/ \mathrm{yr} =[1\times 10^7, 1 \times 10^8, 1 \times 10^9, 1 \times 10^{10}]$ as indicated in the legend. We also show two equilibrium solutions from \textsc{CLOUDY} (with OTS approximation in green and full \texttt{RT} in blue). The thin black line is the \textsc{C}$^2$-\textsc{Ray} result at $t=10^{10}$ yr when using the U-OTS approximation. Note the larger difference between the two curves from \textsc{C}$^2$-\textsc{Ray} after $10^{10}$ yrs compared to the difference of the two curves from \textsc{CLOUDY} and the good agreement between \textsc{C}$^2$-\textsc{Ray} using OTS with \textsc{CLOUDY} using OTS.
\label{ctest1b_plot}
}
\end{center}
\end{figure*}

In the second test, \texttt{TEST1 B}, we use the on the spot approximation as described in Section \ref{ots_expl} and otherwise set the same parameters as in the previous test. We show the ionization profiles in Fig.~\ref{ctest1b_plot} together with two results from \textsc{CLOUDY}, one using the OTS approximation (green dashed line) and the other using full radiative transfer of recombination photons (blue dashed line). As is apparent in Fig.~\ref{ctest1b_plot}, the agreement between the \textsc{C}$^2$-\textsc{Ray} and the OTS solution of \textsc{CLOUDY} is good for those distances where the equilibrium solution has been reached ($r < 12$~kpc at $t=10^{10}$ yr, just as in \texttt{TEST1 A}). The general pattern of evolution of the ionization fractions is quite similar to the one in \texttt{TEST1 A}, although the OTS approximation does of course change the detailed values of the fractions.

The \textsc{CLOUDY} solution using full radiative transfer of recombination photons can be used to evaluate the validity of the OTS approximation for this particular test problem. Inspection of Fig.~\ref{ctest1b_plot} shows that the error introduced by using on the spot approximation is barely noticeable, but larger for hydrogen and largest for the $\mathrm{H I}$  fraction inside the $\mathrm{H II}$ region. Closer inspection shows that the neutral hydrogen profile inside the $\mathrm{H II}$ region, up to a neutral fraction of about 1\%, follows the curve from the $\alpha^{\textrm{A}}$ recombination coefficient from \texttt{TEST1 A}. This is not surprising since most of the recombination photons from this highly ionized region will not be absorbed \em on the spot\em, but escape to larger distances.

To gauge the importance of applying the coupled OTS rather than the popular U-OTS (i.e.\ using $\alpha^B$ rates, see Sect. \ref{no_coupling_eq}), we also plot the ionization profiles after $10^{10}$ yr for the latter approach (thin black lines). It can be seen that the differences between the OTS and U-OTS results are larger than between the OTS approximation and the full radiative transfer results, when comparing the (almost) equilibrium solutions. From this we conclude that although more complicated to implement, the coupled OTS approximation should be the preferred approach.

Since \textsc{C}$^2$-\textsc{Ray} cannot (now) deal with the diffuse photons in another way than using the OTS approximation, we are not able to investigate the effects of using this approximation during the growth of the H~II region. \citet{2011MNRAS.411.1678C} compare results from what we call the U-OTS approximation with full radiative transfer of diffuse photons using their code \textsc{RADAMESH}. They find rather large effects at times when the equilibrium solution has not yet been reached. This warrants further investigation, probably best pursued in manner similar to I+06, involving results from multiple codes.

\subsection{Test-suite 2: A power law source}

In a second set of tests, we use UV-photon emitting sources with a power-law spectrum, $L(\nu) \propto \nu^{-\beta}$, implying that the number of photons goes as $f(\nu) \propto \nu^{-(\beta+1)}$. For these we considered the two cases $\beta=1$ and $\beta=2$. We present the results for simulations with temperature evolution using $T_{\mathrm{ini}}=10^2$ K. We only use the full OTS approximation, apply the secondary ionizations and choose the number of sub-bins in bin 2 and 3 to be $(n_2 , n_3) =(26,20)$. The remaining parameters are as in \texttt{TEST 1}.

We found a timestep of $\Delta t =10^5$ yr to be sufficient to obtain convergence in the temperature evolution. Choosing a timestep of the order of the ionization time of the first cell ($\Delta t =10^3$ yr) gave temperature results which for cells close to the source were only different by at most 7\%. For the parameters of this test, the optical depth at the ionization threshold for hydrogen for one cell is $\tau\sim 3.2$, in between the two cases tested in Appendix~\ref{sec:time_step_dep} and so this result for the timestep is consistent with those.

\begin{figure*}
\begin{center}
\includegraphics[width=15cm]{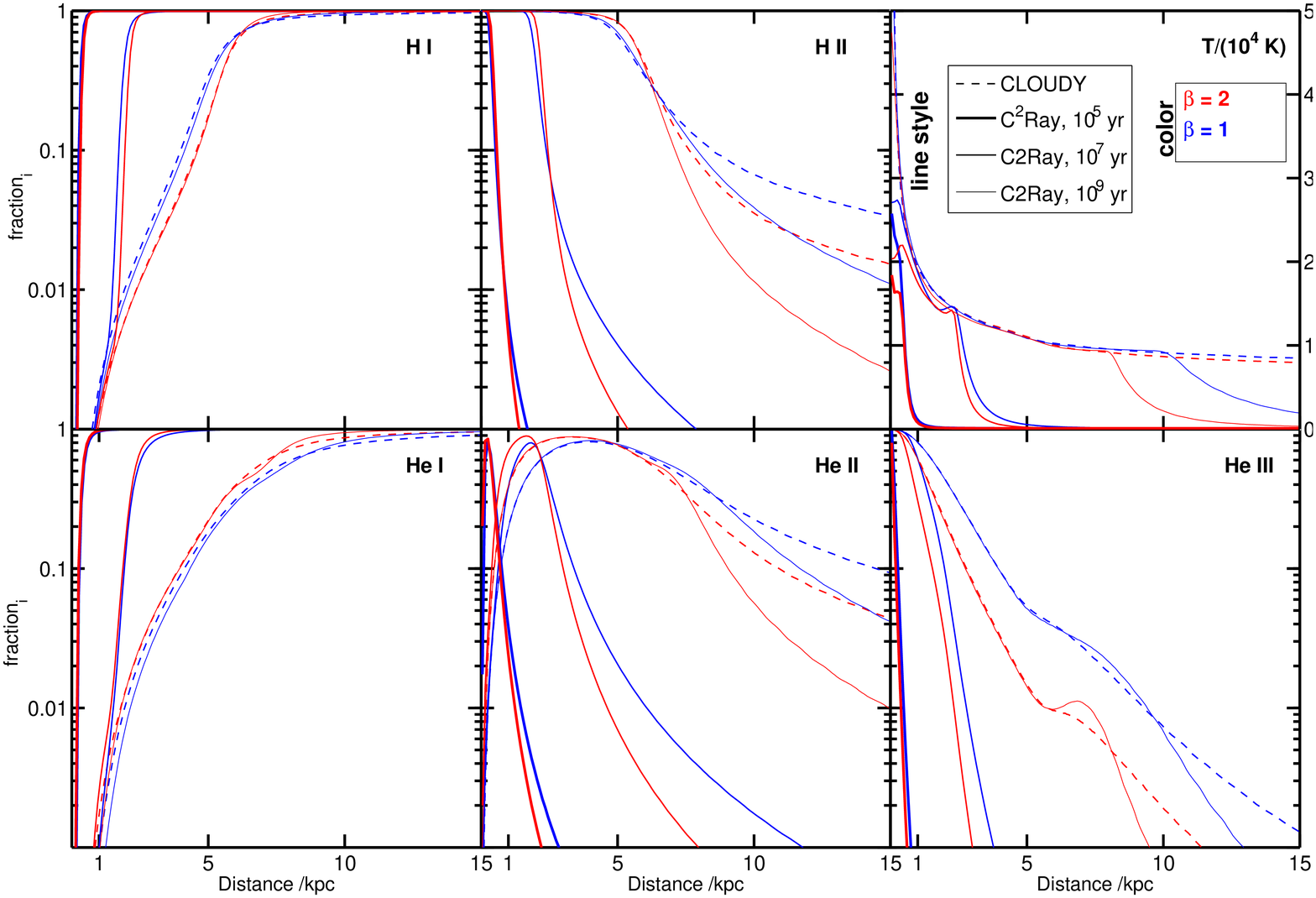}
\caption{Results from \texttt{TEST 2}. Fractions of neutral/ionized hydrogen, neutral/single and double ionized helium and temperature as a function of distance to source for a source with $\beta=2$ (red) and $\beta=1$ (blue) with temperature evolution and secondary ionizations for \textsc{C}$^2$-\textsc{Ray} after $t$/yr $=10^5,\,10^7$ and $10^{9}$ (decreasing line thickness). For comparison, we also include the equilibrium solutions of \textsc{CLOUDY} (dashed lines).
\label{ctest2a_plot}
}
\end{center}
\end{figure*}

We show the profiles for $\mathrm{H I/ H II}$ and $\mathrm{He I/ He II/ He III}$ and $T$ for three times and the two different power law spectra in Fig.~\ref{ctest2a_plot}. We also plot the equilibrium solutions from \textsc{CLOUDY} with OTS for those two spectra. We find that at $t=10^{9}$~yr the \textsc{C}$^2$\textsc{-Ray} results are close to the \textsc{CLOUDY} ones for those distances where the equilibrium solution has been reached, although not as close as for the results of \texttt{TEST 1}. These somewhat larger differences with the \textsc{CLOUDY} results cannot be explained by effects from the temperature calculation since we found comparable discrepancies when imposing a constant temperature. However, overall the match is still reasonable.

The time evolution shows that the two values of $\beta$ produce similar results, with initially quite steep profiles for the ionization fractions for all times and a relatively steep front in the temperature evolution even for the quite hard $\beta=1$ spectrum at the relatively late time of $t=10^7$ yr. Comparing the $\beta=1$ and $\beta=2$ results it can be noticed that the former gives a higher degree of hydrogen and helium ionization outside the front than the latter. Further it can be seen that the residual neutral hydrogen fraction inside the front is higher for the results with $\beta=1$. Just as in TEST1, the bumps in the $\mathrm{He III}$ fraction are transient phenomena, although even the equilibrium solution shows a slope change around a distance of $\sim 6$~kpc.

\citet{2007MNRAS.380.1369T} reported the $\mathrm{H II}$ front to be trailing behind the $\mathrm{He III}$ front for spectra with $\beta<1.8$. However, even in our $\beta=1$ results we find the $\mathrm{H II}$ fraction to be always larger than the $\mathrm{He III}$ fraction. We were able to obtain (double) front crossings when using the $\alpha^A$ recombination rates (no OTS) and disabling the secondary ionizations. However, the locations where the $\mathrm{He III}$  fraction was found to be larger than the $\mathrm{H II}$ fraction were in a limited area in the ionization fraction-time space, roughly in the interval [0.1, $10^8$ yr] to [0.01, $10^9$ yr]. We therefore conclude that trailing $\mathrm{H II}$ fronts are at best a marginal effect.

In order to evaluate the effects of secondary ionizations we also ran \texttt{TEST 2} without them. We found that this led to changes  larger than the differences we found between our results and the equilibrium solution of \textsc{CLOUDY}. We therefore conclude that it is worth to include secondary ionizations in the calculation.

\subsection{Cosmological tests} 
In this section, we test the 3D-version of the extended \textsc{C}$^2$\textsc{-Ray} on two cosmological density fields.
The first test is a larger scale cosmological test without temperature evolution to evaluate the effect of helium at a fixed temperature in a simulation with sources with a soft spectrum. The second test is a rather small cosmological volume but includes temperature evolution: we redo the cosmological test-problem with multiple sources from I+06 (their test4) to test the effect of helium on the heating and on the hydrogen ionization fraction field. 

\subsubsection{TEST 3: Effect of helium on the morphology of the hydrogen ionization fraction field during EoR without temperature evolution for stellar type sources}
\label{sec:cosmo_test}
\begin{figure*} 
\begin{center}
 \includegraphics[width=12cm]{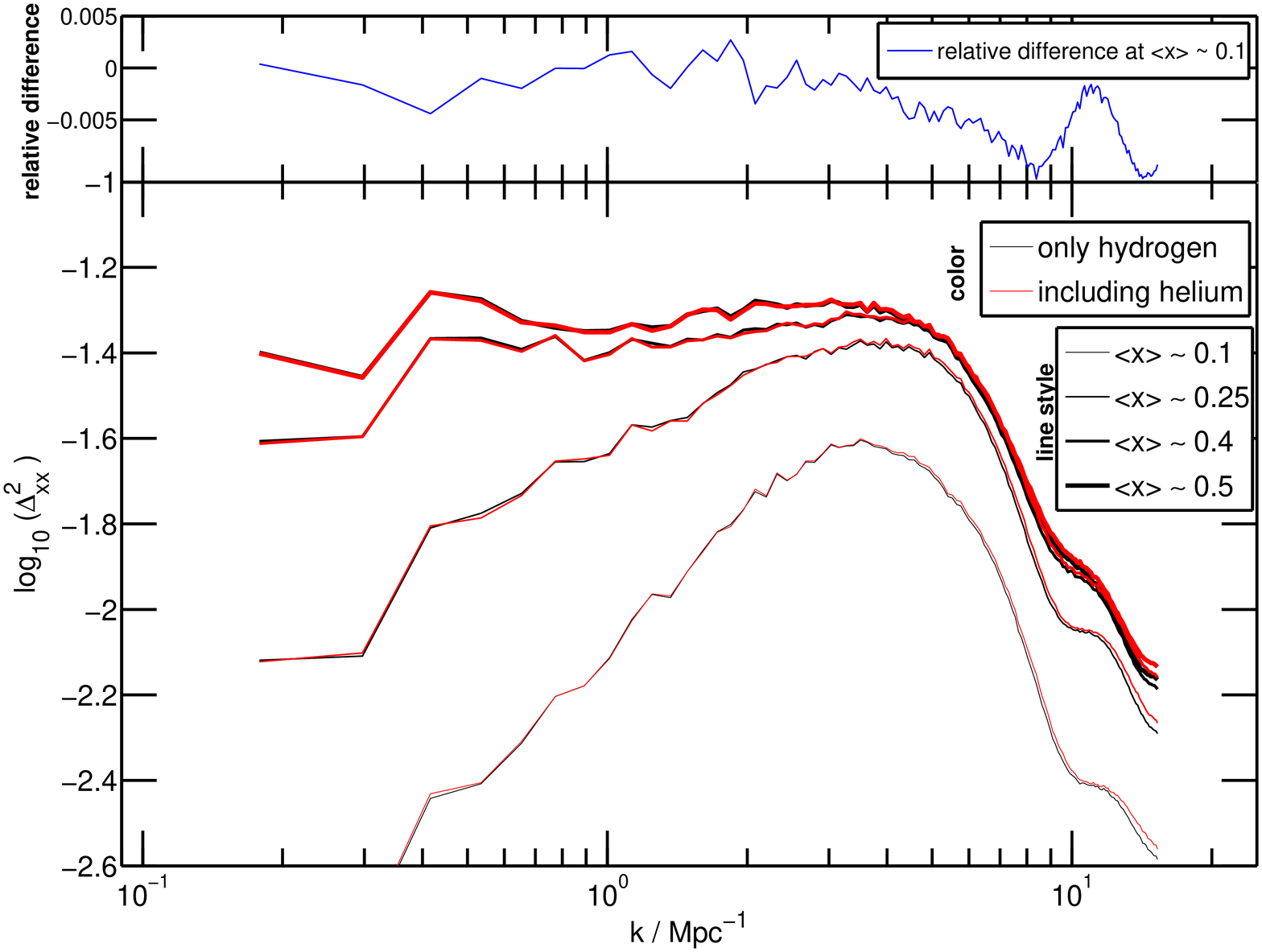}
\caption{Results from \texttt{TEST 3}. Power spectrum of the ionized fraction $x_{\rm{H}^+}$ of simulation 53Mpc\_g8.7\_130S without (black lines) and with (red lines) helium included at four different global (mass averaged) hydrogen ionization fractions $\left< x \right>$ as indicated in the legend. The differences are small: As can be seen in the top panel, for $\left< x \right> \sim 0.1$, the relative difference $\frac{log10\left(\Delta_{xx}^2(H)\right)-log10\left(\Delta_{xx}^2(H+He)\right)}{log10\left(\Delta_{xx}^2(H+He))\right)}$are below 1 \%.
\label{powerspec}
}
\end{center}
\end{figure*}

Many cosmological reionization simulations only include hydrogen and implicitly assume that helium is singly ionized everywhere where hydrogen is ionized. The used hydrogen number density is therefore equal to the total number density. In this section, we test if the morphology of H~II regions in a reionization simulation with stellar sources changes if helium is included. For this comparison, we use simulation 53Mpc\_g8.7\_130S from \citet{2011MNRAS.413.1353F} and \citet{2011arXiv1107.4772I}. The electron scattering optical depth produced by this simulation, $\tau_{\rm{es}}=0.083 $,  is consistent with the 1--$\sigma$ range allowed by the seven year WMAP results, $\tau_\mathrm{es}=0.088 \pm 0.015$ \citep{2011ApJS..192...18K}. In this simulation, the number of ionizing photons produced by a dark matter halo of mass $M$ is defined through
\beq
\dot{N}_\gamma=g_\gamma\frac{M\Omega_b}{10 \Omega_m m_p}\,, 
\eeq
where $\dot{N}_\gamma$ is the number of ionizing photons emitted per
Myr, $\Omega_b=0.044$, $\Omega_m=0.27$ and $m_p$ is the proton mass \citep[This equation included incorrectly a $\mu$ in][ equation 1]{2011MNRAS.413.1353F}.  Massive halos are assigned an efficiency $g_{\gamma}=8.7$ while low mass sources have an efficiency of $g_{\gamma}=130$ and are suppressed in regions where the ionization fraction (of hydrogen) is higher than 10\%. To evaluate the effect of helium on the morphology of H~II regions we use the dimensionless power spectrum of the H~II fraction $\Delta^2_{xx}$ In Fig.~\ref{powerspec} we show the power spectra for this simulation without (black) and with (red) helium included. It can be seen that the power spectra are almost identical. We also show the relative difference, defined as $\frac{log10\left(\Delta_{xx}^2(H)\right)-log10\left(\Delta_{xx}^2(H+He)\right)}{log10\left(\Delta_{xx}^2(H+He))\right)}$ a global ionization fraction of $\left<x\right> \sim 0.1$. The relative error is everywhere below 1\%. This is similar for the other global ionization fractions. Therefore we conclude that the simplification of only using hydrogen for reionization simulations with only stellar-type sources is legitimate. 

\subsubsection{TEST 4: Multiple sources in a small cosmological density field}
This test was fully described in I+06 and has subsequently been used in many papers on radiative transfer methods \citep[e.g.][]{2011MNRAS.411.1678C, 2011MNRAS.415.3731P, 2011MNRAS.412.1943P}. Here we present results for this test including a cosmological helium abundance. We want to use this to illustrate the effect the presence of helium has on this test problem, as well as to compare to the original \textsc{C}$^2$-\textsc{Ray} results from I+06 which used the simpler heating rate calculation, see Sect. \ref{sec:heating}. Here we only summarize the most important aspects of the test setup and refer the reader to I+06 for details: The density field is a snapshot at redshift $z \approx 8.85$. The simulation box has a side length of $0.5/h$ comoving Mpc and the radiative transfer grid consists of $128^3$ uniform cells. The box boundaries are transmissive. The 16 most massive halos in the box constitute the 16 sources. They have a constant photon output during the course of the simulation, the joint ionizing photon rate of all 16 sources is $3.29 \times 10^{53}$ ionizing photons per second. All sources have a black body spectrum with an effective temperature $T_{\rm eff}=100\,000$ K. The initial gas temperature in all cells is $T_{\rm ini} =100$ K. All conditions are as in test4 of I+06 except that our simulation has a helium abundance of $n_{\rm He}/n=0.074$. 

\begin{figure*} 
\begin{center}

 \includegraphics[width=15cm]{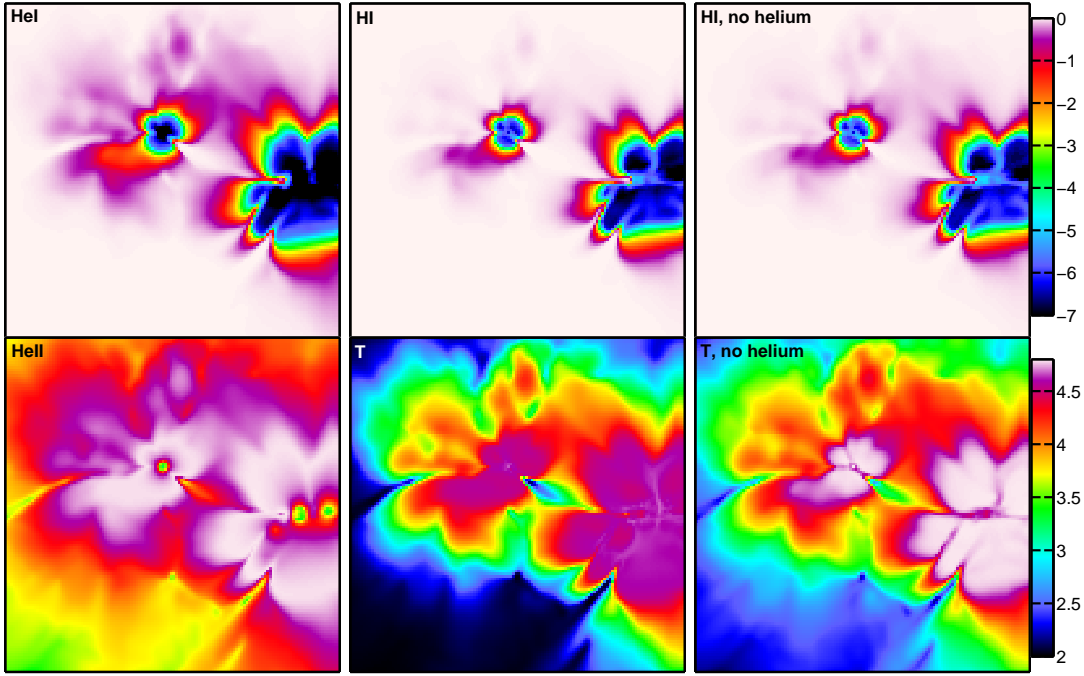}
 \caption{Resuls from \texttt{TEST 4}. Slices through the center of the simulation box of TEST4  I+06 after 0.5 Myr. Upper panels from left to right: HeI, HI, HI from the original \textsc{C}$^2$\textsc{-Ray} (for details see text); lower panels from left to right: HeII, temperature, temperature from the original \textsc{C}$^2$\textsc{-Ray}. The color coding for H and He is logarithmic as indicated by the upper color scale, the temperature is in linear scale as indicated by the lower color scale on the right hand side.} 
\label{fig:TEST4}
\end{center}
\end{figure*}

In Fig.~\ref{fig:TEST4} we show slices through the center of the simulation volume, just as shown in I+06. We show the HI, HeI and HeII fractions (in a logarithmic colour scale) and the temperature (in a linear colour scale) together with the original \textsc{C}$^2$\textsc{-Ray} results presented in I+06 (figures 31 and 32) after 0.05 Myr of evolution. Before describing the visible differences in the results, we need to point out two additional (apart from the helium) important differences which mostly affect the heating: As described briefly in Sect. \ref{sec:heating}, the multi-frequency implementation of the heating forces us to use a timestep close to the ionization time scale. The original \textsc{C}$^2$\textsc{-Ray} implementation used in this test did not have this restriction since it used the constant heating per photo-ionization approach. The \textsc{C}$^2$\textsc{-Ray} results presented in I+06 (right hand panels of  Fig.~\ref{fig:TEST4}) used a timestep $\Delta t = 0.001$ Myr. We now use a timestep $\Delta t = 0.00025$ Myr which was chosen on the basis of convergence studies. 

As can be seen in Fig.~\ref{fig:TEST4}, the effect of using the multi-frequency heating instead of the constant energy- per-ionization heating is a lower temperature inside the HII region. Partly, this is also due to helium since the higher ionization energy of helium results in slightly less energetic photons. However, it can be seen that high density filaments close to the sources are in fact warmer than the less denser regions between them, while it was vice versa in the original implementation. This can be explained as follows: While in the low density region close to sources mainly low energy photons with a higher ionization cross-section are absorbed, in the dense filaments, higher energy photons are absorbed which deposit more energy in the gas. However, dense knots in the filaments which do not host sources, still act as shields against the ionizing and heating radiation, resulting in embedded cold neutral regions. Another obvious difference is the less extended heating front outside the H~II region. This is solely due to the inclusion of helium and its contribution to the optical depth, as tests without helium have shown.  

For the hydrogen ionization fraction field, we note that although the temperature inside the HII region is lower than in the original simulation, the ionization fractions are similar. This is due to the recombination photons from helium. In general, the differences in hydrogen ionization fraction are very small everywhere. Due to the rather high effective temperature, $T_{\rm eff}=100 \, 000$ K, a considerable amount of photons capable of doubly ionizing helium is produced. The resulting He~III regions can be seen as holes in the He~II fraction in the lower left panel of Fig.~\ref{fig:TEST4}. 

The average temperatures inside the H~II regions are now (with the multi-frequency treatment of the heating) more similar to the temperatures from \textsc{CRASH} \citep{2003MNRAS.345..379M} in I+06 and results of \textsc{TRAPHIC} presented in \citet{2011MNRAS.412.1943P}. However, if those results also show higher temperatures for dense filaments inside the ionization front is not clear. 

Temperatures outside the ionization fronts cannot be compared since the results shown here are obtained with helium. The inclusion of helium effectively prevents a preheating ahead of the ionization front. 

\section{Conclusions}
\label{section:conclusions}
We presented an extension of the radiative transfer and photo-ionization code \textsc{C}$^2$\textsc{Ray}, first introduced in \citet{methodpaper} as a method for calculating non-equilibrium hydrogen photo-ionization. The new version treats the transfer of ionizing photons in step with time-dependent photo-ionization of both hydrogen and helium using the full OTS approximation, multi-frequency ionization and heating, as well as secondary ionizations. We described in detail the new elements to \textsc{C}$^2$-\textsc{Ray}, such as the linearized solution of the set of hydrogen and helium rate equations using the coupled OTS treatment, and the calculation of the multi-frequency ionization and heating rates for both hydrogen and helium.

We validated our implementation of the various ionization and heating processes, including the OTS approximation and secondary ionizations, by comparing to results from the photo-ionization equilibrium code \textsc{CLOUDY}. We validated our time-dependent solutions through convergence studies, which also provide us with timestep constraints. We confirmed that the new version of \textsc{C}$^2$-\textsc{Ray} retains the property of being able to calculate ionization fractions with an accuracy of a few percent even for timesteps as large as 0.1 times the recombination time and hard spectra. However, to obtain accurate temperatures, we found the timestep constraints to be more stringent. In the worst case timesteps of about 10\% of the ionization time are needed to correctly determine the temperature. This condition can be relaxed in case the cells are optically thin, or when one is interested in temperatures beyond the recombination time, or when the radiative cooling time is 
of the order of the timestep.

A comparison between results obtained with the OTS approximation and with the often used, simpler U-OTS approximation (perhaps better known as the $\alpha^\mathrm{B}$ approximation), shows that there are significant differences in the results and that these differences are larger than those between the OTS approximation and the case of full radiative transfer of recombination photons. This implies that it is more accurate to use the full OTS approximation. We also verified that secondary ionizations have significant effects for hard enough spectra and should therefore be included.

We used the new version of \textsc{C}$^2$-\textsc{Ray} to re-run a cosmological reionization radiative transfer simulation with stellar-type sources that previously had been simulated with only hydrogen. By comparing the power spectra of the HII fractions, we concluded that the morphologies of HII regions are not different between these two simulations. This confirms that it is correct to assume that for stellar-type sources, the helium ionization follows that of hydrogen.

Finally, we presented the first version with helium for ``Test 4'' from the Cosmological Radiative Transfer Comparison Project (I+06). We find that the inclusion of helium significantly affects the temperature distribution as the heating front is found to be steeper than in the hydrogen only case. This is caused by the larger cross-section of neutral and singly ionized helium at higher photon energies. Helium is therefore stopping hard photons which would otherwise preheat the material far ahead of the the ionization front. This result confirms the original motivation of including helium in \textsc{C}$^2$-\textsc{Ray}, it is an important absorber of EUV and SX photons and should therefore be taken into account when studying hard ionizing spectra.

We are planning to use the new version of \textsc{C}$^2$-\textsc{Ray} to explore the effects of quasar-like sources on reionization, both on the morhphologies of HII regions and the heating of the IGM. However, the new version can be used to study any kind of photo-ionization problem in which substantial amounts of $\mathrm{He III}$ are formed, for example the inner parts of galactic HII regions around O and B stars.

\section*{Acknowledgments} 
MMF is thankful to Anders Jerkstand, Gabriel Altay and Barbara Ercolano for valuable discussions and to Adam Lidz for helpful written communication. 

This study was supported in part by the Swedish Research Council grant 2009-4088. ITI was supported by The Southeast Physics Network (SEPNet) and the Science and Technology Facilities Council grants ST/F002858/1 and ST/I000976/1. PRS was supported by  NSF grants AST-0708176 and AST-1009799, NASA grants NNX07AH09G, NNG04G177G and NNX11AE09G, and Chandra grant SAO TM8-9009X. A significant fraction of the RT simulations  were run on Swedish National Infrastructure for Computing (SNIC) resources at HPC2N (Ume\aa, Sweden) and PDC (Stockholm, Sweden).

\bibliographystyle{mn2e}
\begin{appendix}

\section{Detailed solution to the uncoupled rate equations} 
\label{solution_uncoupled}
For completeness, we present here the solution to the helium rate equations in the uncoupled case (which we introduced as the U-OTS approximation in Sect. \ref{no_coupling_eq}), i.e. the expressions for the eigenvalues, eigenvectors, particular solution vectors and coefficients that can be used in the general solution to Eq.~(\ref{general_problem}) which is given by  Eq.~(\ref{general_solution}). For hydrogen, the values were given in Eq.~(\ref{Eq:H_values}).
\bqa 
\lambda_1^{\mathrm He} &=& \frac{(\mathsf{A}^{\mathrm{He}}_{(11)}+\mathsf{A}^{\mathrm{He}}_{(22)}-\mathrm{S})}{2} \\
\lambda_2^{\mathrm He} &=& \frac{(\mathsf{A}^{\mathrm{He}}_{(11)}+\mathsf{A}^{\mathrm{He}}_{(22)}+\mathrm{S})}{2} \\
\mathbfsf{x}_1^{\mathrm He} &=&\left(\begin{array}{c}    
                                   -\frac{-\mathsf{A}^{\mathrm{He}}_{(11)}+\mathsf{A}^{\mathrm{He}}_{(22)}+\mathrm{S}}{2 \mathsf{A}^{\mathrm{He}}_{21}}\\1
                                    \end{array}\right)\\
\mathbfsf{x}_2^{\mathrm He} &=&\left(\begin{array}{c} 
				   -\frac{-\mathsf{A}^{\mathrm{He}}_{(11)}+\mathsf{A}^{\mathrm{He}}_{(22)}-\mathrm{S}}{2 \mathsf{A}^{\mathrm{He}}_{21}}\\1
                                    \end{array}\right)\\
\mathbf{x}_p^{\mathrm He} &=&          \left(\begin{array}{c}
	\frac{\mathbf{g}(1) \mathsf{A}^{\mathrm{He}}_{(22)}}{-\mathsf{A}^{\mathrm{He}}_{(11)} \mathsf{A}^{\mathrm{He}}_{(22)}+\mathsf{A}^{\mathrm{He}}_{(21)}\mathsf{A}^{\mathrm{He}}_{(12)} }\\
	\frac{-\mathbf{g}(1)\mathsf{A}^{\mathrm{He}}_{(21)}}{-\mathsf{A}^{\mathrm{He}}_{(11)} \mathsf{A}^{\mathrm{He}}_{(22)}+\mathsf{A}^{\mathrm{He}}_{(21)}\mathsf{A}^{\mathrm{He}}_{(12)} }
                                    \end{array}\right)\\
c_1^{\mathrm He} &=&\frac{-\mathrm{R} \, \mathsf{x^{He}_2}(1) + \mathrm{T} }{ \mathsf{x^{He}_1}(1)-\mathsf{x^{He}_2}(1)} \\
c_2^{\mathrm He} &=&\frac{ \mathrm{R} \, \mathsf{x^{He}_1}(1) -\mathrm{T}}  { \mathsf{x^{He}_1}(1)-\mathsf{x^{He}_2}(1)} \, ,
\eqa
where $\mathrm{S}$,  $\mathrm{T}$ and $\mathrm{R}$ are given by:
\bqa
\mathrm{S} &=& \! \sqrt{ (\mathsf{A}^{\mathrm{He}}_{(11)})^2  +
                      (\mathsf{A}^{\mathrm{He}}_{(22)})^2  +
                      4 \mathsf{A}^{\mathrm{He}}_{(12)}  \mathsf{A}^{\mathrm{He}}_{(21)} \!-
                    \! 2 \mathsf{A}^{\mathrm{He}}_{(11)}  \mathsf{A}^{\mathrm{He}}_{(22)}} \\
\mathrm{T} &=& x_0^{\mathrm{He}}(1)-x_p^{\mathrm{He}}(1) \\
\mathrm{R} &=& x_0^{\mathrm{He}}(2)-x_p^{\mathrm{He}}(2) 
\eqa
This solution was also given in \citet{2008MNRAS.386.1931A}.

\citet{1987A&A...174..211S} who introduce the same general problem, Eq.~(\ref{general_problem}), split up the matrix in two 2-state systems and introduced a coupling between the stages which results in a single time dependence for the two stages of helium. This solution was used for example by \citet{1994A&A...289..937F}, \citet{1997ApJS..109..517R}, \citet{1998A&A...331..335M}, \citet{2002A&A...394..901M} and  \citet{2004MNRAS.348..753S}. Here, we include explicitly the two different time dependences (the two different exponential constants), namely the two eigenvalues $\lambda_1^{\mathrm He}$ and $\lambda_2^{\mathrm He}$. Their contribution is weighted differently for the two species of helium, see Eq.~(\ref{general_solution}). This is the mathematically correct solution and, as an extreme example suggests, also makes sense physically: Assume for the ionization rate $ \Gamma_{\rm {HeII}} \lra 0$  and initially, $x_{\rm He III} > 0$, say $x_{\rm He III}=0.5$. In this case, the time evolution of $x_{\rm HeIII}$ should not (directly) depend on the ionization rate $\Gamma_{\rm He I}$, so the time evolution of $x_{\rm HeII}$ and $x_{\rm HeIII}$ should not have the same exponential factor.

The averages of the ionization fractions over one timestep $\Delta t$  can be calculated for both hydrogen and for helium as:
\bq
\left<\mathbf{x} \right> = \frac{\int \mathbf{x}(t) dt} {\int t \, dt}= \sum_{i=1}^3 c_i \mathbfsf{x}_i \,  \mathfrak{A}_i + \mathbf{x}_p \quad \rm{with} \quad \mathfrak{A}_i= \frac{(\rm{e}^{\lambda_i \Delta t}-1)}{\Delta t \, \lambda_i}
\label{average1}
\eq
To avoid floating point precision problems $\mathfrak{A}_i$ is set to 1 explicitly when $\lambda_i \Delta t < 10^{-8}$.

\section{Detailed solution to the coupled rate equations} 
\label{solution_coupled} 
In the case of the OTS approximation, the rate equations for hydrogen and helium are coupled by the recombination photons from the different ionization stages of helium.  The solution to  Eq.~(\ref{general_problem}) with the elements defined as in Eq.~(\ref{coupled_x_g}) and Eq.~(\ref{coupled_A}) is given by Eq.~(\ref{general_solution}) with the following values for the eigenvalues, eigenvectors,  particular solution vector and coefficients $c_i$:
\bqa
\lambda_1 &=& \mathsf{A}_{(11)} \\
\lambda_2 &=& 0.5 (\mathsf{A}_{(33)}  + \mathsf{A}_{(22)} - \mathrm{S})\\
\lambda_3 &=& 0.5 (\mathsf{A}_{(33)}  + \mathsf{A}_{(22)} + \mathrm{S}) 
\eqa
%
\begin{align}
\mathbfsf{x}_1 &=\left(\begin{array}{c} 1 \\ 0 \\ 0 \end{array}\right) \\ 
\mathbfsf{x}_2 &=\left(\begin{array}{c} \frac{-2\mathsf{A}_{(32)} \mathsf{A}_{(13)}  + \mathsf{A}_{(12)} (\mathsf{A}_{(33)}  -\mathsf{A}_{(22)} + \mathrm{S})}{2 \mathsf{A}_{(32)} (\mathsf{A}_{(11)}- \lambda_2)}\\ \frac{ -\mathsf{A}_{(33)}  + \mathsf{A}_{(22)}  - \mathrm{S}  }{2 \mathsf{A}_{(32)} } \\ 1 \end{array}\right) \\
\mathbfsf{x}_3 &=\left(\begin{array}{c} \frac{-2\mathsf{A}_{(32)} \mathsf{A}_{(13)}  + \mathsf{A}_{(12)} (\mathsf{A}_{(33)}  -\mathsf{A}_{(22)} - \mathrm{S} )}{2 \mathsf{A}_{(32)} (\mathsf{A}_{(11)}-\lambda_3)}\\ \frac{ -\mathsf{A}_{(33)}  + \mathsf{A}_{(22)}  + \mathrm{S}  }{2 \mathsf{A}_{(32)} } \\ 1 \end{array}\right)
\end{align}
%
%
\bqa
\mathbf{x}_p=  \left( \begin{array}{c} -\frac{1}{\mathsf{A}_{(11)}}\left( 
  \mathbf{g}(1)+(\mathsf{A}_{(33)} \mathsf{A}_{(12)}- \mathsf{A}_{(32)} \mathsf{A}_{(13)})\mathbf{g}(2) \mathrm{K}
 \right)   \\
 \mathsf{A}_{(33)} \mathbf{g}(2) \mathrm K \\
-\mathsf{A}_{(32)} \mathbf{g}(2) \mathrm K
                        \end{array}\right)
\eqa
\begin{align}
c_1 &= 
\frac{ -2 \mathbf{x_p}(1) \mathrm{S}- \left(\mathrm{R}+(\mathsf{A}_{(33)}- \mathsf{A}_{(22)}) \mathrm{T}  \right)(\mathbfsf{x}_2(1)- \mathbfsf{x}_3(1))}
{2 \mathrm \, S} +  \nonumber\\ 
& \mathbf{x_0}(1)+\frac{\mathrm{T}}{2}\,(\mathbfsf{x}_2(1)+ \mathbfsf{x}_3(1) ) \\
c_2 &= \frac{\mathrm{R}+(\mathsf{A}_{(33)}- \mathsf{A}_{(22)}-\mathrm{S}) \mathrm{T}}{2 \,\mathrm S} \\
c_3 &= -\frac{\mathrm{R}+(\mathsf{A}_{(33)}- \mathsf{A}_{(22)}+\mathrm{S}) \mathrm{T}}{2 \,\rm S} 
\label{ci_coubpled}
\end{align}
Here, the coefficients $\mathrm S$, $\mathrm K$, $\mathrm R$ and $\mathrm T$ are defined as
\bqa
\mathrm{S} &=&\sqrt{\mathsf{A}_{(33)}^2-2\mathsf{A}_{(33)}\mathsf{A}_{(22)}+\mathsf{A}_{(22)}^2+4\mathsf{A}_{(32)}\mathsf{A}_{(23)}} \\
\mathrm{K} &=&1/(\mathsf{A}_{(23)} \mathsf{A}_{(32)} - \mathsf{A}_{(33)} \mathsf{A}_{(22)}) \\
\mathrm{R} &=& 2 \mathsf{A}_{(23)} \left(\mathbf{x}_p(2)-\mathbf{x}_0(2) \right)\\
\mathrm{T} &=& \mathbf{x}_p(3)-\mathbf{x}_0(3)
\eqa
The averages can be calculated with Eq.~(\ref{average1})
where $\lambda_i$, $c_i$, $x_p$ and $\mathbfsf{x}_i$ as introduced in this Appendix.

\section{Ionization cross sections}
\label{cross_sections}

\begin{figure} 
\begin{center}
\includegraphics[width=9cm]{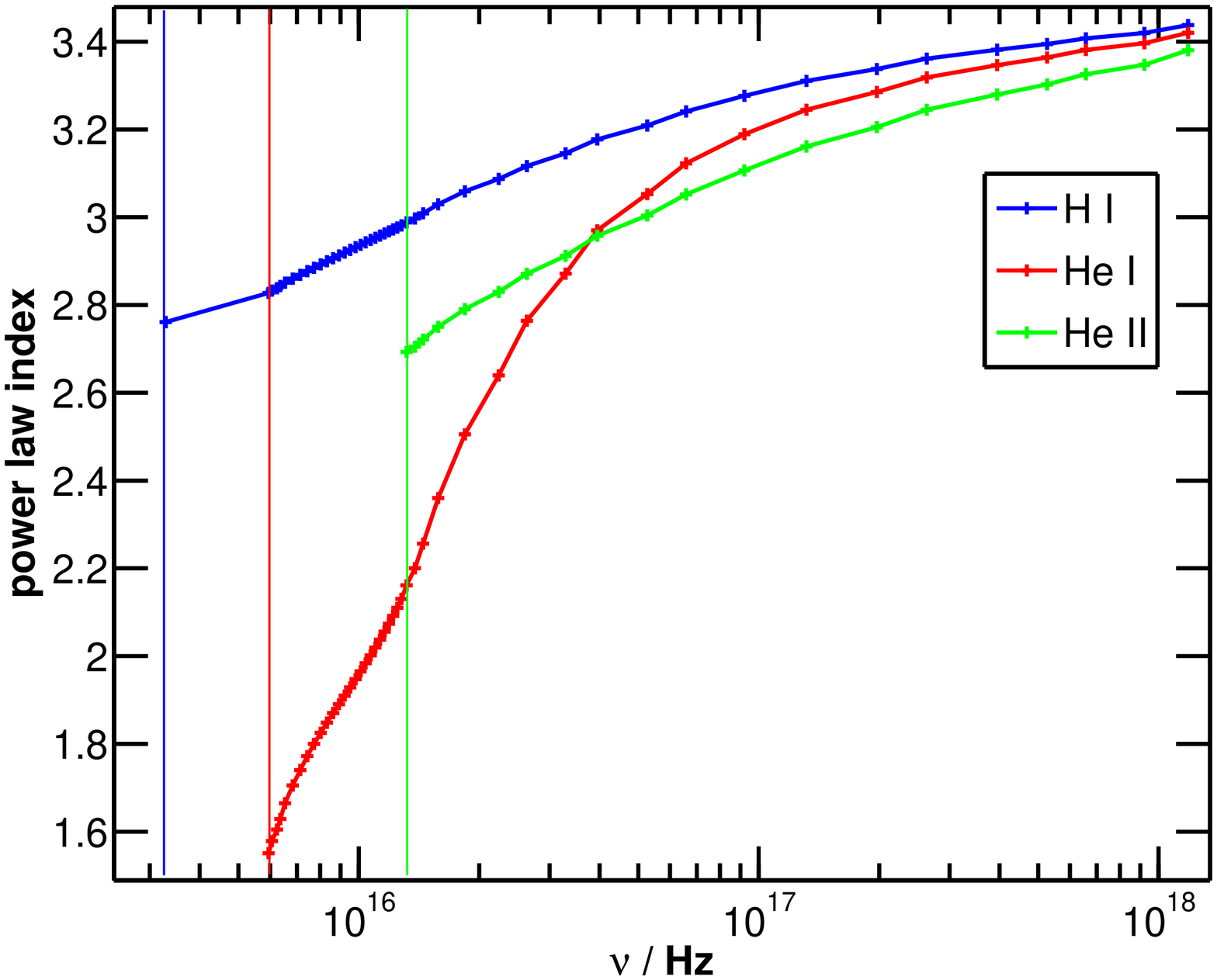}
\caption{Power law indices of $\mathrm{HI}$ (blue), $\mathrm{HeI}$ (red) and $\mathrm{HeII}$ (green) ionization cross-section from the fits to \citet{1996ApJ...465..487V}. The vertical lines indicate the 
threshold frequencies. Note the very different power law indices for $\mathrm{HI}$ and
He~I in frequency bin 2 
and the very similar power law indices for all three species above $10^{17}$ Hz. 
\label{cross_secs2}
}
\end{center}
\end{figure} 

As described in Section \ref{radiation}, within each sub-bin we assume the same frequency dependence 
of the ionization cross sections for all three species. In all sub-bins in bin 2, we 
use the power law indices from our fit to the  neutral helium ionization cross-section 
data from \citet{1996ApJ...465..487V}. In the sub-bins from bin 3, we use the power law indices from 
the fit to the ionized helium ionization cross-section. To fit the power-law, we 
use a linear-least square fit in the $(\log_{10}\nu, \log_{10}\sigma)$ space. 
The power-law indices are presented in figure \ref{cross_secs2} for (1,26,20) 
subbins in bin (1,2,3). 
\section{On the division of photons}
\label{photondivision}
Several different suggestions for the distribution of the ionizing photons between different species have been proposed in the literature. We follow the method 
proposed by A. Lidz (private communication) and \citet{2008MNRAS.386.1931A}, who distribute the photons depending on the ratios of the optical depth. For two species labelled 1 and 2:
\bq
f_1=\frac{\tau_1}{\tau_1+\tau_2},
\label{lidz}
\eq
 where $f_1$ is the fraction of ionizations going into ionizing species 1 with $\tau_1$.
\citet{2004MNRAS.348L..43B} suggested the following recipe
\bq
f_1=\frac{(1-\exp(-\tau_1))\exp(-\tau_2)}{(1-\exp(-\tau_1))\exp(-\tau_2)+
(1-\exp(-\tau_2))\exp(-\tau_1)}
\label{bolton}
\eq
and \citet{2003MNRAS.345..379M} chose
\bq
f_1=\frac{1-\exp(-\tau_1)}{(1-\exp(-\tau_1))+(1-\exp(-\tau_2))}
\label{anto}
\eq

\begin{figure} 
\begin{center}
 \includegraphics[width=8.7cm]{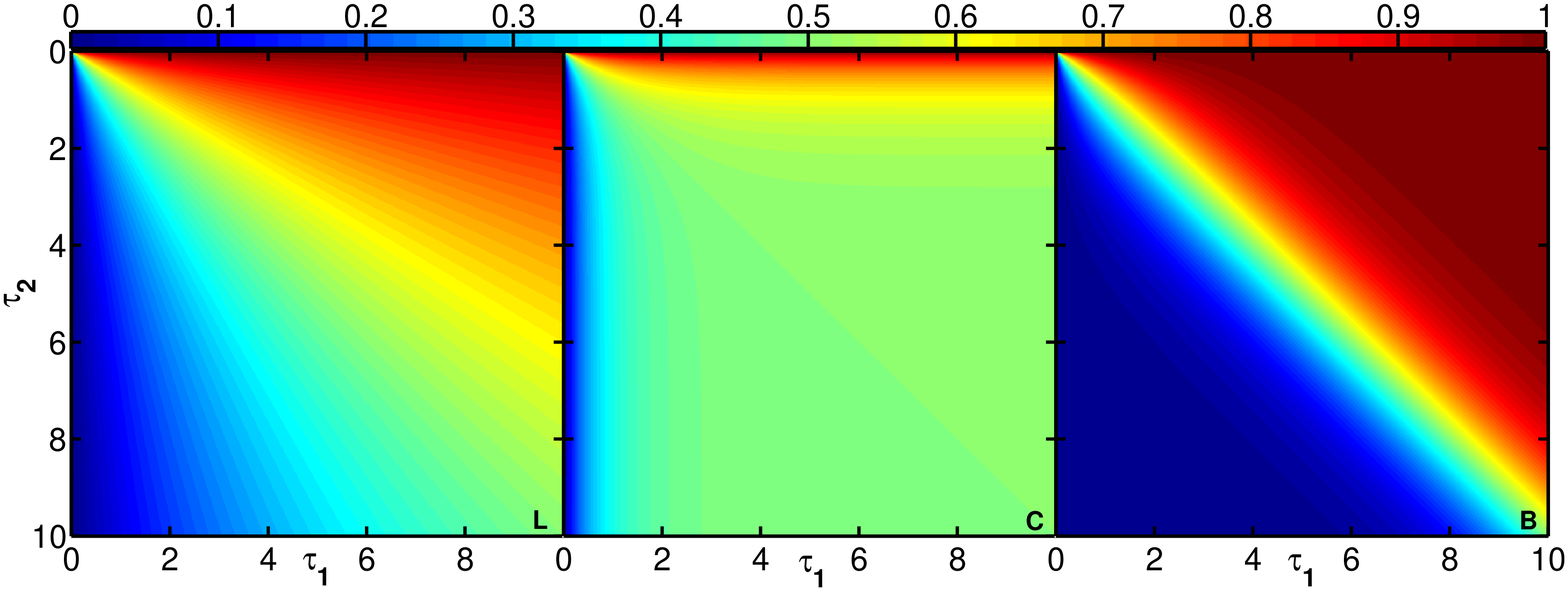}
\caption{For a two species medium, we show the fraction of ionizing photons 
absorbed by species 1  as a function of the optical depth of 
both species using the three different recipes, L (Eq.~\ref{lidz}, left panel), C 
(Eq.~\ref{anto}, middle panel) and B (Eq.~\ref{bolton}, right panel). The fraction 
is color coded as indicated by the color bar, red corresponding to 
$f_1=1$ and blue corresponding to $f_1=0$.
\label{dividing_phots}
}
\end{center}
\end{figure}

By considering the limits of low-optical depth limit and high optical depth limit it becomes clear that Eq.~(\ref{lidz}) is the correct one: For the low optical depth limit, the intensity leaving a cell with optical depth $\tau$ is
$I(\tau)=I_0 \, (1-\tau)$, where $I_0$ is the ingoing intensity. The absorbed fraction, $(I_0-I(\tau))/I_0$ depends therefore linearly on $\tau$. For very high optical depths, the fractions according to Eq.~(\ref{bolton}) and Eq.~(\ref{anto}) become largely independent of the fraction of actual optical depth of the two species: Independent of the exact ratio of the optical depth, Eq.~(\ref{anto}) yields values close to 0.5 for both fractions while the fractions are close to 1 and close to 0 for Eq.~(\ref{bolton}). This cannot be correct as the fractions should still depend on the relative values of $\tau_1$ and $\tau_2$.
This is illustrated in Fig.~\ref{dividing_phots}, where we plot the fraction $f_1$ as a function of optical depth of both species for all three recipes. 

We would like to point out that the latest version of \textsc{CRASH} now also uses Eq.~(\ref{lidz})
to distribute the photo-ionizations over the different species (Maselli, private
communication). 

\section{Varying the number of sub-bins in the frequency bins 2 and 3}
\label{Subbins}

As mentioned in Section \ref{radiation}, the code can use different numbers of sub-bins in frequency bins 2 and 3. As described in Appendix \ref{cross_sections}, the code uses a single power law fit for all species in each sub-bin. 

This has the following disadvantage: In frequency bin 2, for example, a single power law fit to the ionization cross-sections of each species (separately), He I and H I would be a sufficiently good fit and yield power law indices: $s_{\rm H I}=2.91$ and $s_{\rm He I}=1.88$. However, since these are very different numbers, using 1.88 as a power law index over the entire bin 2 for both species, introduces a substantial error in hydrogen optical depth. Therefore, we introduce sub-bins.  

\begin{figure*} 
\begin{center}
 \includegraphics[width=8.7cm]{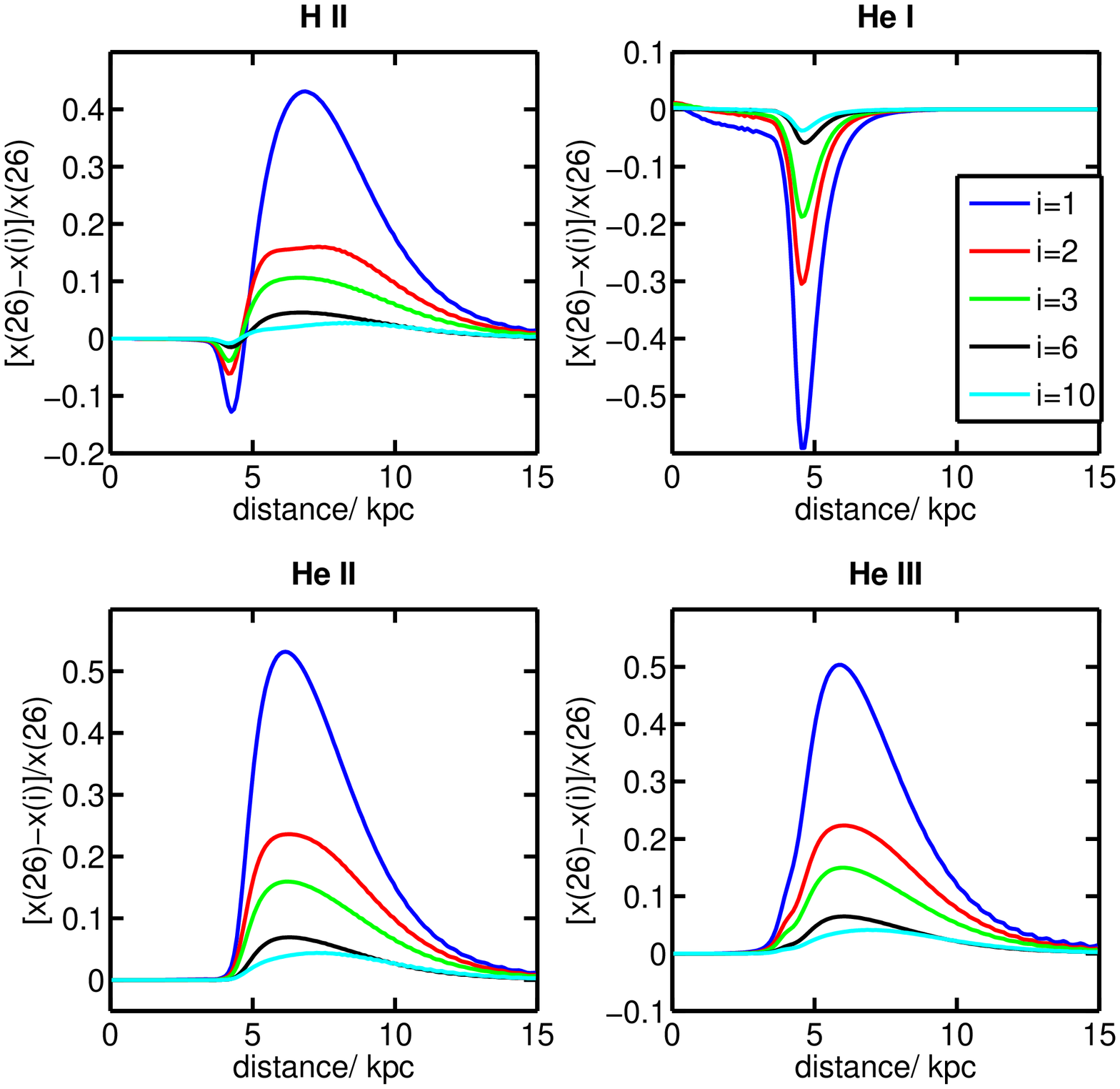}
 \includegraphics[width=8.7cm]{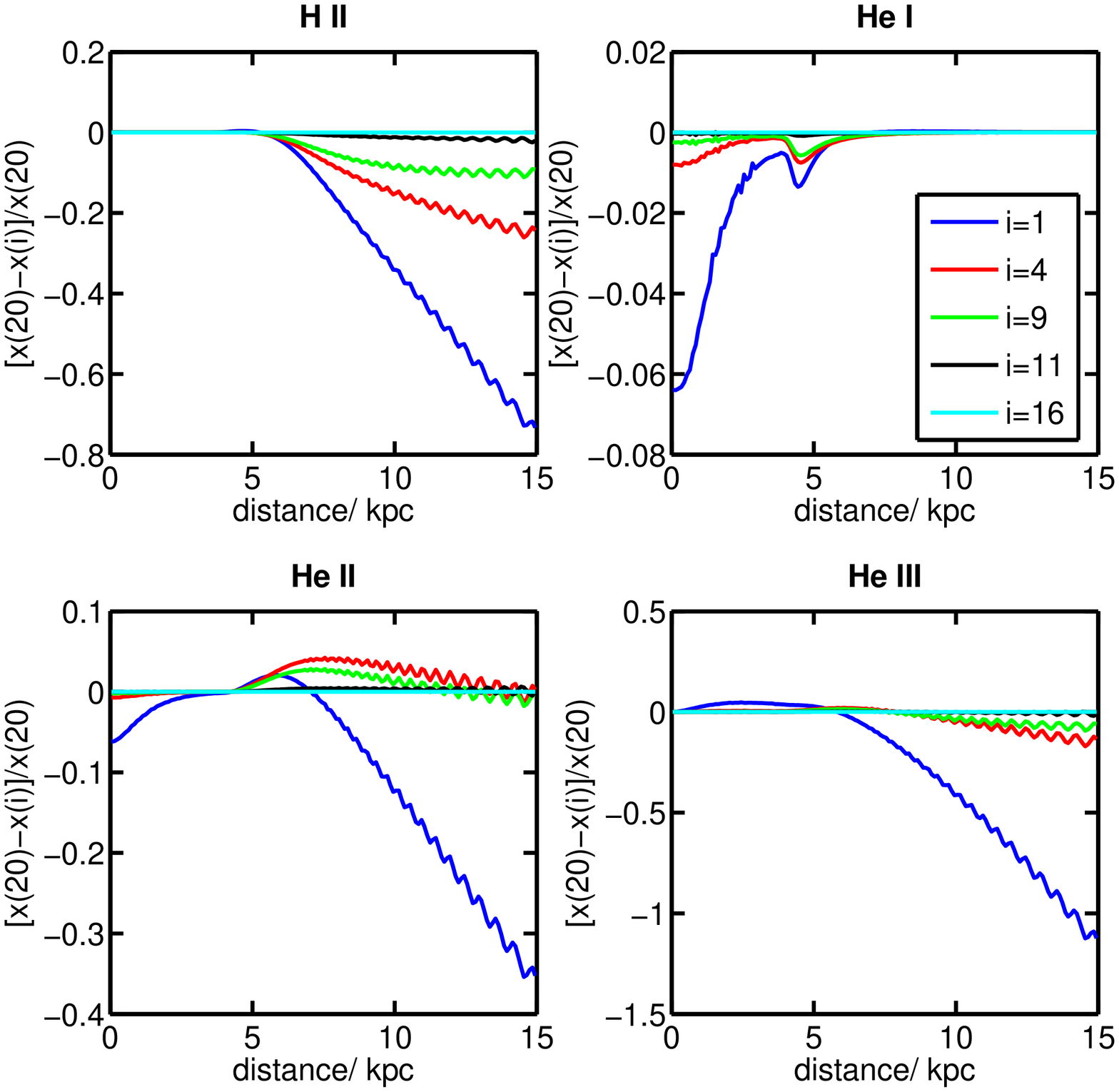}
\caption{Relative difference of $\mathrm{H II}$, $\mathrm{He I}$, $\mathrm{He II}$ and $\mathrm{HeIII}$ fractions. The input parameters are as in \texttt{TEST1B}. In the four panels to the left, the number of sub-bins in frequency-bin 2 is varied according to the legend. In the four panels on the right hand side, the number of sub-bins in frequency-bin 3 is varied according to the legend.  
\label{sub_bin_conv}
}
\end{center}
\end{figure*}

We tested different numbers of sub-bins for bins 2 and 3. For this test, we use the parameters from \texttt{TEST2} with $\beta=0$ in order to have a substantial fraction of ionizing photons in bin 3. We show our results in Fig.~\ref{sub_bin_conv} where the four panels on the left hand side show the relative differences of the ionization fractions (H II, He I, He II and He III) using 1,2,3,6 and 10 sub-bins in bin 2 as compared to using 26 sub-bins. The four panels on the right hand side show the relative difference of the ionization fractions using 1,4,9,11 and 16 sub-bins in bin 3 as compared to using 20 sub-bins. For bin 2, it can be seen that the relative differences using 10 sub-bins compared to 26 is at most at 4 \%. This maximum error is at the ionization front position. Here, the error introduced by using the OTS approximation as opposed to including diffuse photons is most probably higher than that. Therefore, we conclude that using more than 10 sub-bins in bin 2 might not sufficiently improve the results. For bin 3, it can be seen that the relative differences using 11 sub-bins compared to 20 is at most 2 \% which is reached far outside the front in the H~II and He~III fractions. Apart from those locations, the error is well below the percentage level. We therefore conclude that for most applications, 11 sub-bins in bin 2 are sufficient.   

\section{Dependence on timestep}
\label{sec:time_step_dep}
As was pointed out in M+06, the approximation of using time-averaged values for the ionization states in the  calculation of the ionization rates to avoid the need of small timesteps is strictly only valid in the case of negligible contribution from collisional ionizations and recombinations. For those processes it does matter at which time during the timestep, they occur. 

Given the added complexity due to the inclusion of helium, the coupled OTS approximation and the multi-frequency photo-heating, we present in this Appendix convergence tests for the timestep. We first consider the convergence of the ionization fractions at constant temperature and then the convergence of the temperature evolution.  

\begin{figure*} 
\begin{center}
 \includegraphics[width=12cm]{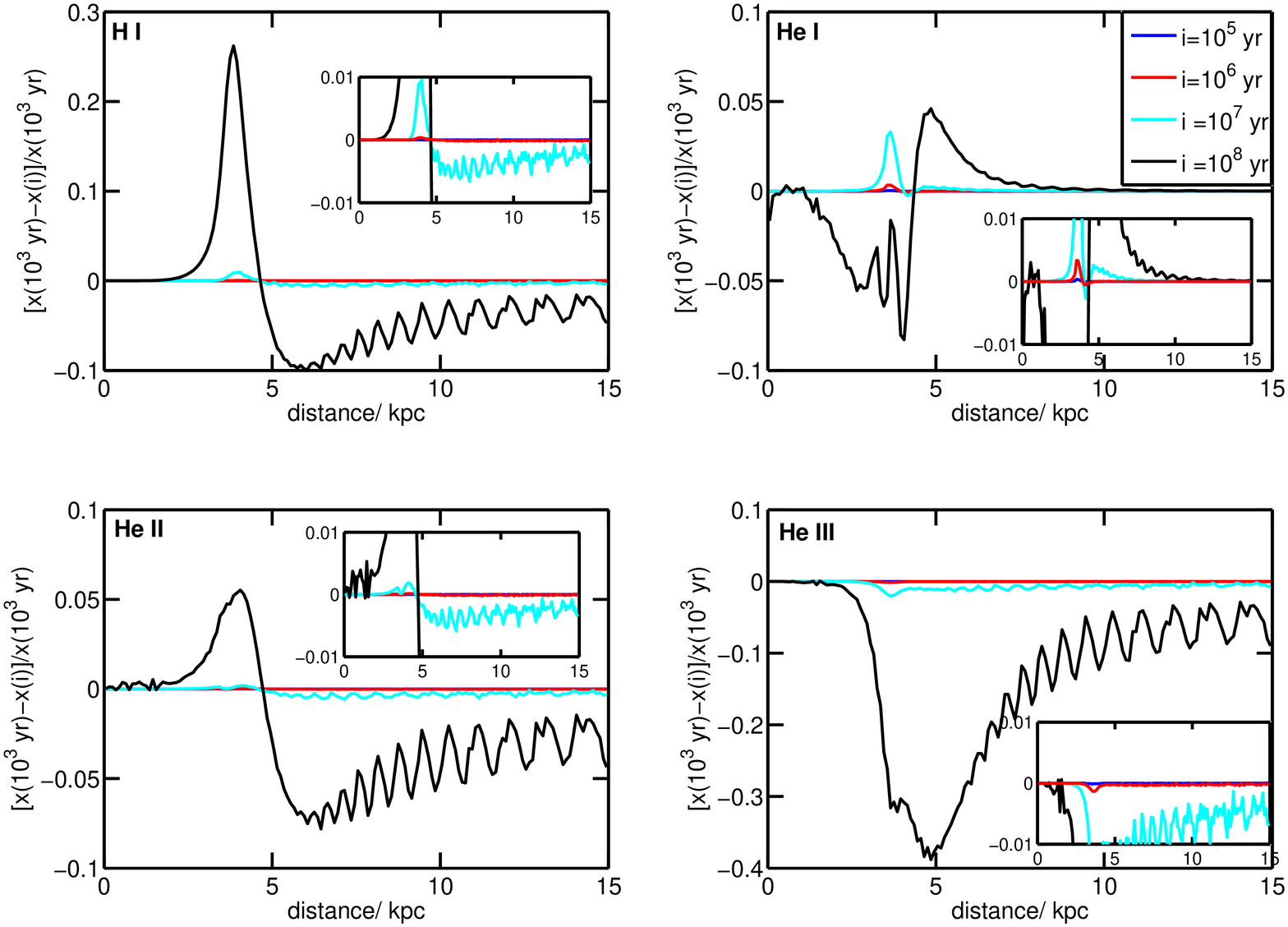}
\caption{Relative difference of $\mathrm{H II}$, $\mathrm{He I}$, $\mathrm{He II}$ and $\mathrm{He III}$ fractions at $t=10^8$ yr. The timestep $\Delta t$ was varied according to the legend. The comparison is made against a timestep that correspond roughly to the ionization time of the first cell, $\Delta t= 10^3$ yr.  The input parameters are as in \texttt{TEST2} with $\beta=1$ but without temperature evolution. The insets shows the $\pm$ 1\% regions. 
\label{delta_t_conv}
}
\end{center}
\end{figure*}

We test the effect on the ionization fractions of varying the timestep 5 orders of magnitude for a source with a power-law spectrum with power law index $\beta=1$. Fig.~\ref{delta_t_conv} shows the relative error for a test with the same parameters as in \texttt{TEST2} at $t=10^8$ yr (this corresponds roughly to the recombination time scale) using $\Delta t=10^5, 10^6, 10^7$ and $10^8$ yr, compared to $\Delta t =10^3$.  The latter corresponds roughly to the ionization time for the first cell. As can be seen, even for the large timestep $\Delta t=10^7$, the maximum error is 3\% and for $\Delta t=10^6$ yr, the error is everywhere well below the percent level. We therefore conclude that $\Delta t \leq 0.1 t_{\rm rec}$ is a sufficient timestep criterion for accurate ionization calculations.

Next, we test the temperature evolution dependence on the timestep. Here, we show the results of two one dimensional simulations, one with optically thin cells ($\tau \sim 0.6$) and one with moderately optically thick cells ($\tau \sim 60$). Both have recombination timescales $t_{\rm rec} \sim 10^5$ yr. The ionization time scale for the first cell for the simulation with optical thin cells is $t_{\rm ion} \sim 5 \times 10^{-3}$ yr and the the moderately optically thick case has $t_{\rm ion} \sim 0.5$ yr.  The parameters for these tests are $n_H=0.926$ cm$^{-3}$, $n_{\rm He}=0.074$ cm$^{-3}$  ,$\dot{N}_{\gamma} = 10^{46}$ ($10^{48}$), $\Delta r= 10^{12}$ ($10^{14}$) km, $T_{\rm eff}=100000$ K, $T_{\rm ini}=100$ K for the optically thin (moderately thick) case. We use ($n_2$,$n_3$)=(26,20) frequency sub-bins. 

\begin{figure*} 
\begin{center}
 \includegraphics[width=15cm]{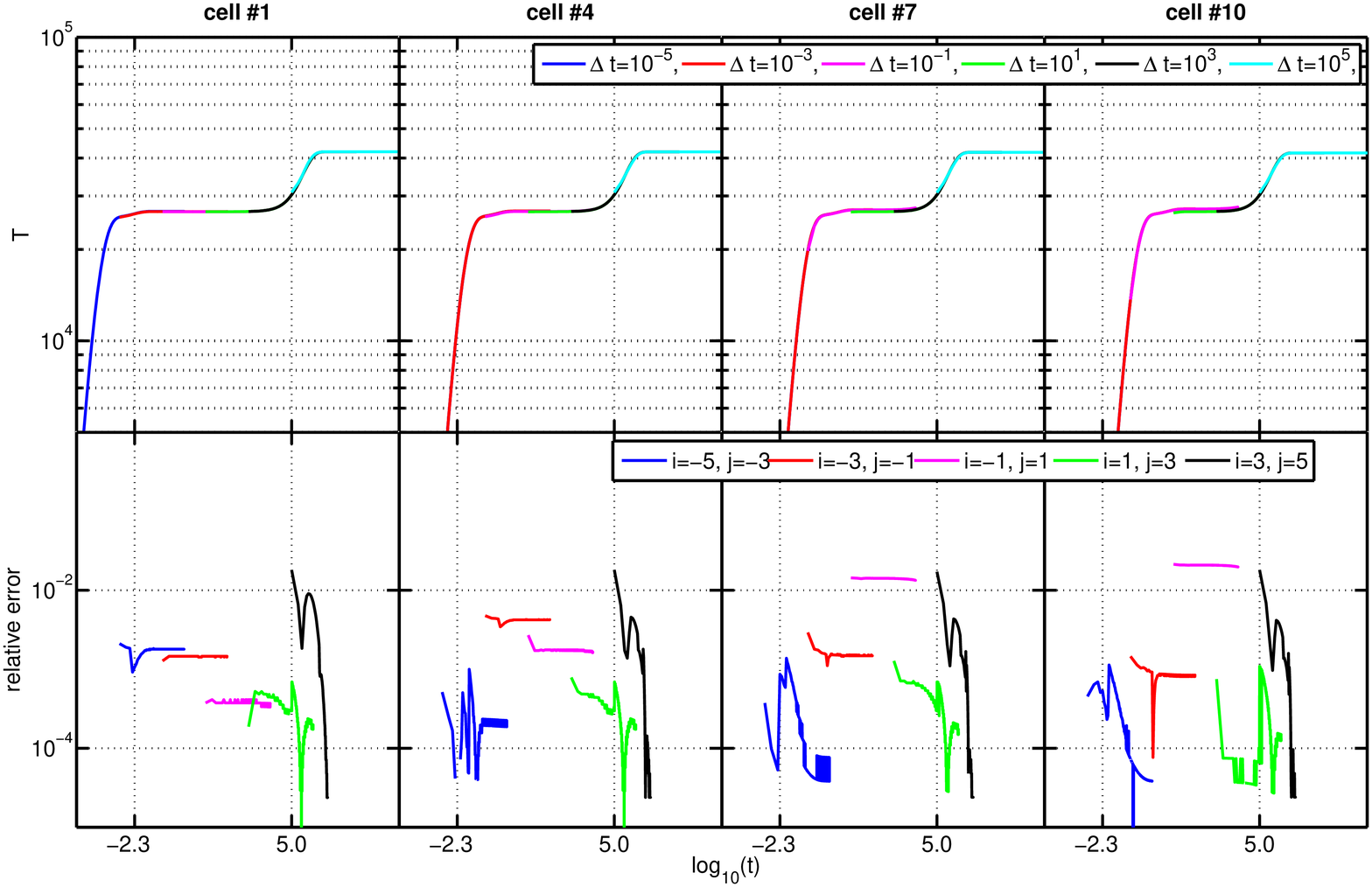}
 \includegraphics[width=15cm]{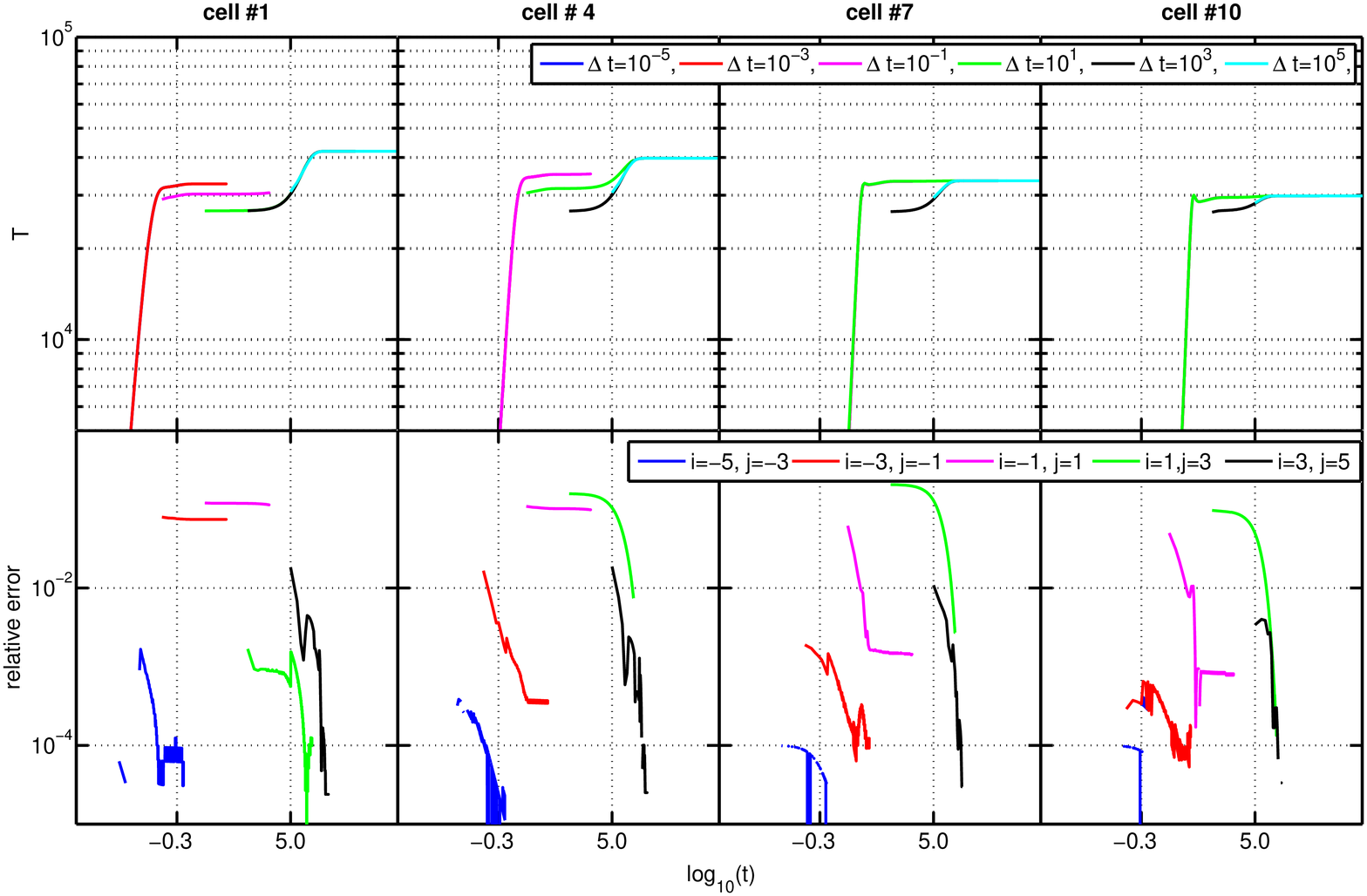} 
\caption{Upper panel: Temporal evolution of the temperature for 4 optically thin ($\tau \sim 0.6$) cells (panels left to right correspond to cells 1, 4, 7 and 10). Lower panel: the same for moderately optical thick cells ($\tau \sim 60$). We show results for 6 different timesteps according to the legend. All times are in years. With each timestep $\Delta t$ we evolved the box for a time corresponding to $10^5 \times \Delta t$. We also show the relative error between the results using two consecutive timestep sizes, according to the legend: relative error = $\frac{T(\Delta t=10^{i} {\rm yr})- T(\Delta t=10^{j} {\rm yr})}{T(\Delta t= 10^{i}{\rm yr})}$. Indicated on the abscissa (and by the dotted vertical grid lines) are the recombination time scale and the ionization time scale (of the first cell). Note that the relative error for the optically thin case is larger for cells further away of the source while the error decreases with distance to the source for the optically thick case.
\label{delta_t_conv_2}
}
\end{center}
\end{figure*}

In Fig.~\ref{delta_t_conv_2} we present the results for these two tests. The form was inspired by that of test0 in I+06 and test1 in \citet{2011MNRAS.412.1943P}. For four cells of our computational grid we show the temperature evolution for six different choices of timestep: $\Delta t$/yr $= 10^{-5}$, $10^{-3}$, $10^{-1}$, $10^{1}$, $10^{3}$ and  $10^{5}$. As we evolve each case only for $10^5$ timesteps, the results of each simulation only overlaps with two others. When two curves overlap only the one with the longest timestep is shown. The lower rows show the relative error between two subsequent choices of timestep.

We see from this figure that while timesteps larger than $t_{\rm ion}$ still yield reasonable good results for the optically thin case, this is not true for the optically thick case. For the optically thin case, the error increases with distance to the source. This is due to an underestimation of the preheating of cells further away from the source. For those the heating rate as function of time is already a substantial fraction of its maximum before the front reaches the cell. This dependence on the optical depth between a cell and the source precludes ``local'' fixes based on the cell properties and evolution.

For the optically thick case, timesteps of the order of ionization timescale result in errors of the order of 8\% (for the first cell) compared to timescales several (2 and 4) orders of magnitude smaller than the ionization timescale. Although this error is larger, it actually decreases with distance, mostly because less preheating occurs at larger distances. 

If one is only interested in time-scales larger than the recombination time-scale, the errors are small in both the optically thick and optically thin case. This is most probably because the initial spike in the heating rate becomes a less important contribution to the total photon energy input in each cell.  

\section{Recombination- cooling and collisional ionization rates}
\label{cooling_recom_rates}

\textbf{RECOMBINATION} For the hydrogen recombination rates, $\alpha_{\rm HII}^A$ and  $\alpha_{\rm HII}^B$ and for the helium recombination rates, $\alpha_{\rm HeIII}^A$ and $\alpha_{\rm HeIII}^B$, we are using the fitting formula from \citet{1997MNRAS.292...27H}, henceforth HG97. These are also good fits to the data from \citet{1994MNRAS.268..109H} (accurate to 1.5 \% in the temperature range prensented there, 10 K to $10^7$ K) even if they were fitted to the slightly older data from \citet{1992ApJ...387...95F}. 

The recombination coefficients $\alpha_{\rm HeII}^A$ and $\alpha_{\rm HeII}^B$ above $T=7 \times 10^4$K are dominated by dielectronic recombination. We include it above $T=1.5 \times 10^4$K according to the fitting formula from \citet{1973A&A....25..137A} as an extra contribution to the recombination rates. Below $T=9 \times 10^3$K, the fitting formula for $\alpha_{\rm HII}^A$ and $\alpha_{\rm HII}^B$ from HG97 provide a good fit (accurate to 6 \% from $T=10$ K to $9 \times 10^3$ K) to the data  for $\alpha_{\rm HeII}^A$ and $\alpha_{\rm HeII}^B$ from \citet{1998MNRAS.297.1073H}. Above $T= 9 \times 10^3$ K, we are using the fits for $\alpha_{\rm HeII}^A$ and $\alpha_{\rm HeII}^B$ from  HG97 which provide fits accurate to 6 \% up to $T=2.5 \times 10^4$K, the highest temperature \citet{1998MNRAS.297.1073H} provides data for.

For the recombination to $n=2$, $\alpha_{\rm HeIII}^2$, we fit the data from \citet{2006agna.book.....O} as $\alpha_{\rm HeIII}^2(T)= 3.4 \times 10^{-13} (T/10^4 K)^{-0.6}$.
\newline

\noindent \textbf{COLLISIONAL IONIZATION} For the collisional ionization coefficients $C_{\rm HI}$, $C_{\rm HeI}$ and $C_{\rm HeII}$ we use the fitting formulas from HG97 in the temperature range $2 \times 10^5$ K $ < T < 10^7$ K where $C_{\rm HI}$  ($C_{\rm HeI}$) fit the data from \citet{1987ephh.book.....J} to an accuracy of 12 \% (5 \%). Below $2 \times 10^5$ we use the fitting formula from \citet{1970PhDT.........2C} which is simpler and slightly more accurate than HG97 for lower temperatures.  
\newline

\noindent \textbf{COOLING} For cooling coefficients for $ {\rm H II}$ and ${\rm He III}$ (free-free + recombination cooling) we interpolate the data from \citet{1994MNRAS.268..109H}. For the ${\rm H I}$ cooling coefficient we include collisional excitation cooling and use the line strength from \citet{1983MNRAS.202P..15A}. For the ${\rm He II}$ cooling coefficient we include collisional excitation, collisional ionization and dielectronic recombination cooling from HG97 and B recombination and free-free cooling from \citet{1998MNRAS.297.1073H}. For ${\rm He I}$ we only include collisional ionization (from HG97) since its collisional excitation is negligible and the rate proportional to $n_e^2$ \citep[c.f.][]{1981MNRAS.197..553B}.

In case of cosmological simulations, we include Compton cooling according to \citet{1987ApJ...318...32S} as well as cooling due to cosmological expansion. 

\end{appendix}

\label{lastpage}
\end{document}